\documentclass{article}
\usepackage{graphicx}
\usepackage[centertags]{amsmath}
\usepackage{amsfonts}
\usepackage{amssymb}
\usepackage{amsthm}

\title{Mechanics of Systems of Affine Bodies. Geometric Foundations and Applications in Dynamics of Structured Media}

\author{J. J. S\l awianowski, V. Kovalchuk, A. Martens,\\ 
B. Go\l ubowska and E. E. Ro\.zko\\
Institute of Fundamental Technological Research,\\
Polish Academy of Sciences,\\
$5^{\rm B}$, Pawi\'{n}skiego str., 02-106 Warsaw, Poland\\
e-mails: jslawian@ippt.gov.pl, vkoval@ippt.gov.pl,\\ 
amartens@ippt.gov.pl, bgolub@ippt.gov.pl, erozko@ippt.gov.pl}

\begin{document}

\maketitle
\begin{abstract}
Discussed are geometric structures underlying analytical mechanics of systems of affine bodies. Presented is detailed algebraic and geometric analysis of concepts like mutual deformation tensors and their invariants. Problems of affine invariance and of its interplay with the usual Euclidean invariance are reviewed. This analysis was motivated by mechanics of affine (homogeneously deformable) bodies, nevertheless, it is also relevant for the theory of unconstrained continua and discrete media. Postulated are some models where the dynamics of elastic vibrations is encoded not only in potential energy (sometimes even not at all) but also (sometimes first of all) in appropriately chosen models of kinetic energy (metric tensor on the configuration space), like in Maupertuis principle. Physically the models may be applied in structured discrete media, molecular crystals, fullerens, and even in description of astrophysical objects. Continuous limit of our affine-multibody theory is expected to provide a new class of micromorphic media.

\end{abstract}
  
\section*{Introduction}

In our former papers, cf. e.g. \cite{acta,all04,all05} and references therein, we discussed in various aspects the mechanics of affinely-rigid bodies, i.e., bodies rigid in the sense of affine geometry. Being interesting in themselves even from the purely academic point of view of analytical mechanics, they were also thought on as a useful model of internal and collective degrees of freedom of various realistic systems. Everything done in the mentioned papers concerned a single affine body (see Fig. 1). 
\begin{center}
\includegraphics[scale=0.25]{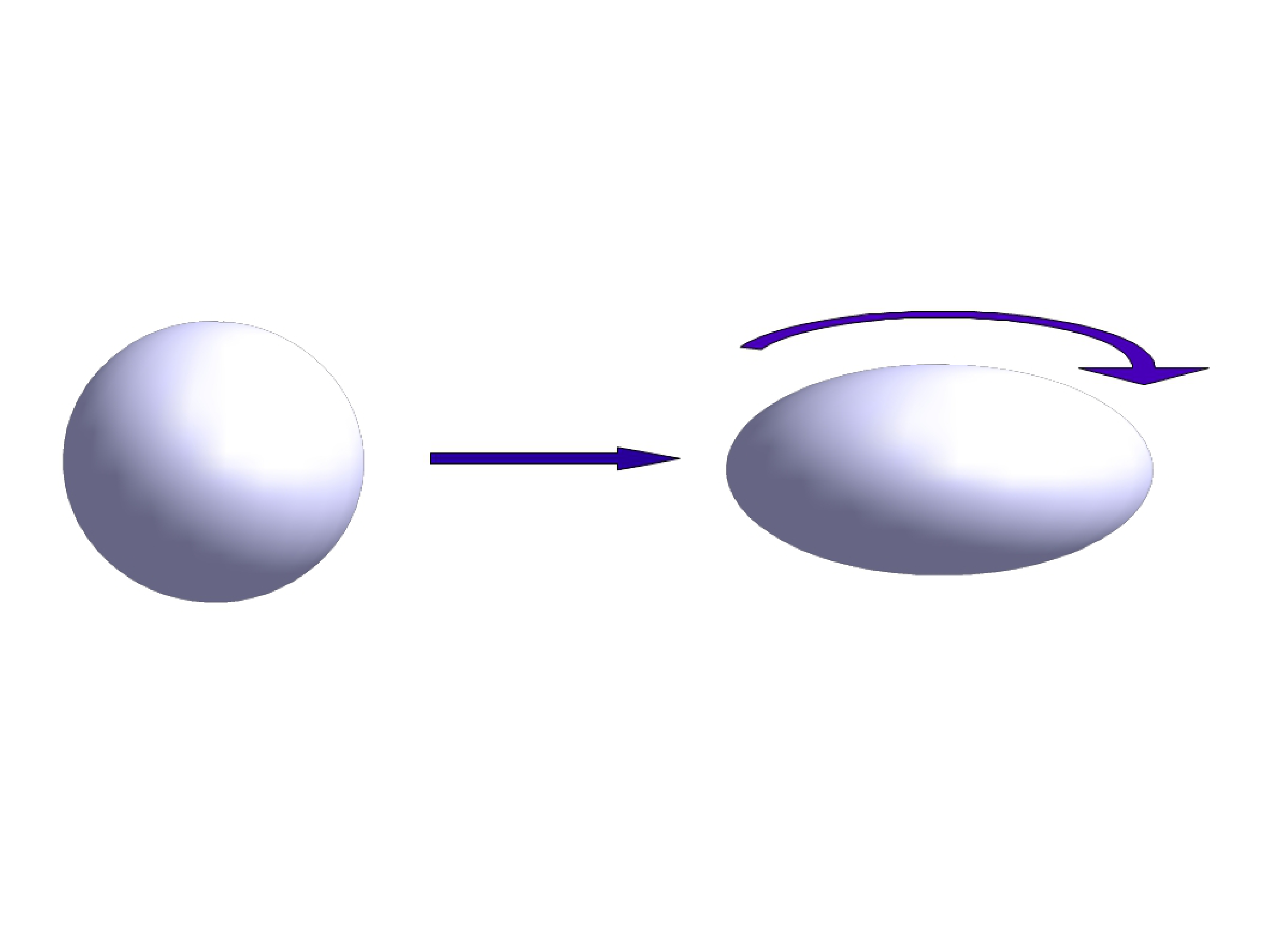}

{\rm Fig. 1. Affine body, degrees of freedom.}
\end{center}
It is interesting, however, to investigate the mechanics of systems of such bodies. Certain quite realistic physical applications are obvious. For example, in various problems of molecular dynamics the most relevant degrees of freedom are spatial translations, rotations and homogeneous deformations. These are so-to-speak the leading modes of motion of a single molecule. In molecular crystals one deals with aggregates, systems of such objects (see Fig. 2). 
\begin{center}
\includegraphics[scale=0.25]{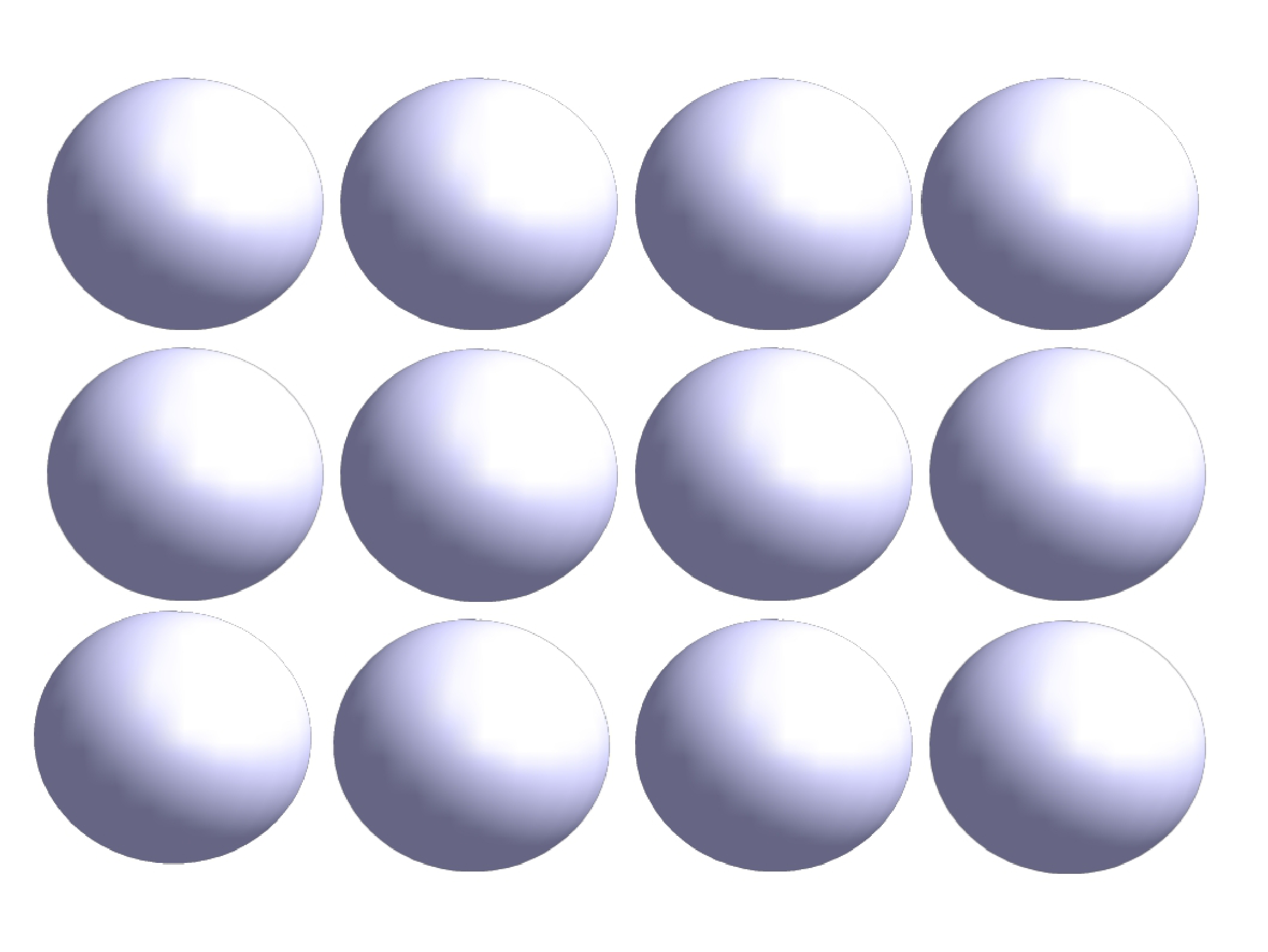}

{\rm Fig. 2. Molecular crystal or granular medium as a system of spheres or spheroids.}
\end{center}
Atoms in a single molecule are strongly coupled to each other, whereas the mutual coupling between separate molecules is relatively weak. So, molecules have a well-defined individuality and the model of a system of mutually interacting affinely-rigid bodies looks physically reasonable and viable. In the long-wave approximation one obtains asymptotically the micromorphic continuum in the sense of Eringen \cite{Capr_89,Erin_64,Erin_68}. This is the continuum of infinitesimal affinely-rigid bodies, consisting of material points with attached linear frames representing internal degrees of freedom, i.e., microrotations and microdeformations if using the terms accepted in theory of micropolar and micromorphic media. In the droplet model of nuclei the collective phenomena are also sometimes described in terms of affine degrees of freedom \cite{B-M_2}; the nuclei-nuclei interaction also may be interpreted in terms of systems of affine grains-droplets. There are vibrating stars and other astrophysical objects. As shown by Riemann, Dedekind, Chandrasekhar and others, their relevant, leading, degrees of freedom in a good approximation are also ruled by the affine group. So, for example, a double star, at least when its constituents are placed relatively near to each other, is a system of mutually interacting affine bodies \begin{center}
\includegraphics[scale=0.25]{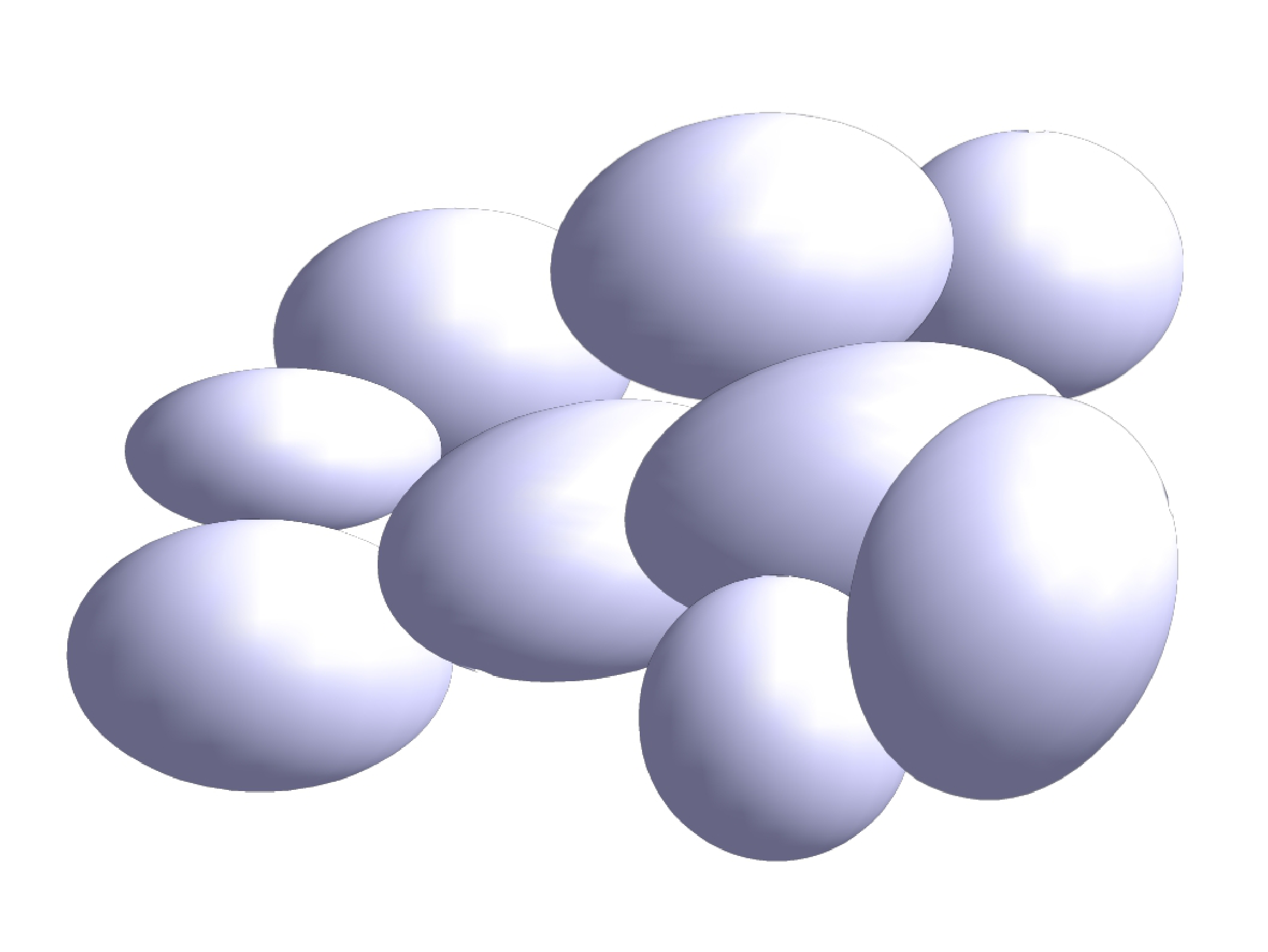}

{\rm Fig. 3. Densely packed grains: microcontinua?}
\end{center}
In all those examples (molecular crystals are perhaps most suggestive, but amorphous media also provide a good example, cf. Fig. 3) the mutual displacements result in stresses and the mutual rotations and deformations lead to hyperstresses (see Fig. 4). 
There is also another application, strongly related to the finite element method. As known, this method was elaborated primarily for the computational, numerical purposes. The material body is "triangulated" like in geodesy and represented as an aggregate of simplexes which are so small that in a good approximation they may be considered as homogeneously deformable, i.e., as affine bodies (according to the very idea of differential calculus, every sufficiently regular configuration may be in sufficiently small regions approximated by affine ones). The boundary surface of the body may be then literally triangulated into two-dimensional, i.e., planar affine bodies, the surface finite elements. In traditional applications all these simplexes were used as elements of the mesh for the purely numerical solving of partial differential equations of elasticity (or viscoelasticity, plasticity, etc.). But nowadays another approach became fashionable, namely one combining in an inventive way rigorous analytical, qualitative and numerical methods \cite{Rub_1,Rub_2}. By introducing the simplex mesh one represents the body as an aggregate of mutually interacting small affine bodies. 
\begin{center}
\includegraphics[scale=0.30]{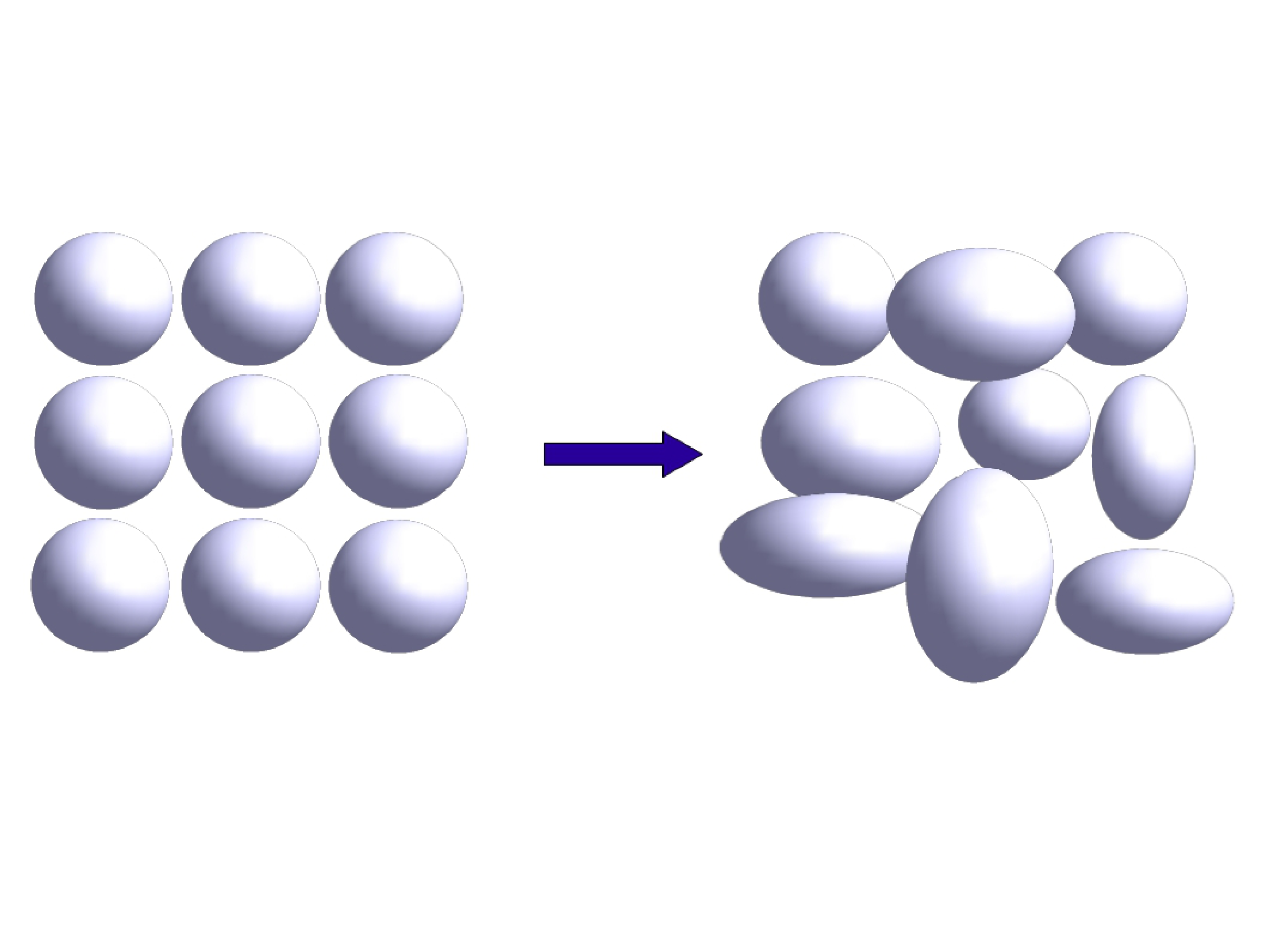}

{\rm Fig. 4. Grains of the regular lattice become translated, rotated and deformed.}
\end{center}
The continuous dynamical problem based on partial differential equations is then approximated by a discrete one, ruled by a finite system of ordinary differential equations with time as the only independent variable \cite{La-Ra-Sha}. Obviously, this reduction to analytical mechanics with a finite number of degrees of freedom is an approximation, the better one, the mesh is more dense and inventive. But one can think about combining the well-elaborated methods of the theory of finite-dimensional dynamical systems, both rigorously analytical and qualitative ones, with the numerical techniques based on the discretization. The procedure becomes then more reliable and the danger of computer artefacts is remarkably smaller than when dealing with the purely mesh-based numerical treatment of dynamical partial equations \cite{Rub_1,Rub_3}. There is also another problem on a more fundamental level. Namely, one can think that in certain situations the simplicial-triangulation modes of motion may be more adequate than, e.g., the usual phononic modes corresponding to the Fourier expansion onto plane waves. In any case this is some reasonable alternative approach. Incidentally, when we go with the size of simplicial elements to the nano-scale, the quantum treatment becomes necessary, just like in the traditional phonon-modes description.

Even if assumed models of mutual interactions (described, e.g., by binary potential) are formally simple, it is rather a rule than exception that, both computationally and qualitatively, the mechanics of a system of material points is much more complicated than the mechanics of a single material point moving in some fixed external field. The situation becomes even incomparatively more difficult when our elementary entities are no longer material points but structured objects, e.g., rigid bodies or affinely-rigid bodies. This fact is well known even in relatively simple systems of mutually coupled gyroscopes. The nontrivial geometry and topology of "internal" degrees of freedom brings about some completely new problems and difficulties, even on the very fundamental level of invariance principle. Although the objects move in the flat affine or Euclidean space (here we consider only such a situation as realistic), the full configuration space is non-Euclidean. When we deal with systems, this leads to really serious complications like difficulty if not impossibility at all of defining correctly the internal motion "as a whole" (some "average" internal motion of the system) and the "relative" internal motions. This is similar and as a matter of fact geometrically strongly related to the impossibility of the correct definition of the centre of mass for a system of material points in a curved (Riemannian) space.

There is also another important novelty when we pass from mechanics of a single object moving in a Lie group as the configuration space (Euclidean, affine, orthogonal, linear) to a system, i.e., when the configuration space becomes the Cartesian product of several copies of such a group. For many reasons, both computationally practical and fundamentally theoretical, particularly interesting are dynamical models invariant under the group underlying geometry of degrees of freedom. If one deals with a single object and its configuration space is just a Lie group, then the only invariant models are geodetic ones, i.e., with vanishing (constant) potentials; Lagrangian is then simply given by the kinetic energy (its metric is left- or right-invariant under regular translations). The reason is obviously that the only functions on group invariant under regular translations (it does not matter left or right) are constant ones. However, if one deals with the system of such objects, i.e., if the configuration space is some Cartesian-product-power of the underlying group, then of course there exist nontrivial potentials, more generally --- nontrivial forces, invariant under this group (left or right acting). This is a very simple algebraic fact. If $G$ is the mentioned group and we consider, e.g., a two-body system with the configuration space $G^{2}=G\times G$, then obviously any function $F:G\times G\rightarrow Y$ of the form
\begin{equation}\label{eq.I.1}
F\left(g_{1},g_{2}\right)=f\left(g^{-1}_{1}g_{2}\right)
\end{equation}
is invariant under simultaneous left regular translations of its arguments:
\begin{equation}\label{eq.I.2}
\left(g_{1},g_{2}\right)\mapsto \left(gg_{1},gg_{2}\right),\qquad g\in G,
\end{equation}
where $Y$ is an arbitrary set and $f:G\rightarrow Y$ is an arbitrary $Y$-valued function on $G$. In particular, $F$, $f$ may be real-valued potential functions.

Similarly, $F$ given by
\begin{equation}\label{eq.I.3}
F\left(g_{1},g_{2}\right)=f\left(g_{2}g^{-1}_{1}\right)
\end{equation}
is invariant under simultaneous right regular translations:
\begin{equation}\label{eq.I.4}
\left(g_{1},g_{2}\right)\mapsto \left(g_{1}g,g_{2}g\right),\qquad g\in G.
\end{equation}
For some special choices of $f$ one can obtain "potentials" invariant simultaneously both under left and right regular translations:
\begin{equation}\label{eq.I.5}
\left(g_{1},g_{2}\right)\mapsto \left(kg_{1}l,kg_{2}l\right),\qquad k,l\in G.
\end{equation}

An interesting example is provided by characters of finite-dimensional representations of $G$. Let $T:G\rightarrow{\rm GL}(X)$ be such a representation; $X$ is some finite-dimensional linear space and
\begin{equation}\label{eq.I.6}
T\left(g_{1}g_{2}\right)=T\left(g_{1}\right)T\left(g_{2}\right),\qquad T(e)={\rm id}_{X},
\end{equation}
where $e\in G$ is the neutral element (identity) of the group $G$. Let $f:G\rightarrow \mathbb{R}$ be a character of $T$:
\begin{equation}\label{eq.I.7}
f\left(g\right):={\rm Tr}\ T\left(g\right).
\end{equation}
It is obvious that
\begin{equation}\label{eq.I.8}
f\left(km\right)=f\left(mk\right),\qquad F\left(k,m\right)=F\left(m,k\right)
\end{equation}
and the function $F$ is invariant both under (\ref{eq.I.2}) and (\ref{eq.I.4}).

All these concepts become particularly intuitive when we consider systems of rigid bodies, when $G={\rm SO}\left(3,\mathbb{R}\right)$, the group of rotations in $\mathbb{R}^{3}$, or its "quantum-spinorial" universal covering $G={\rm SU}\left(2\right)$. The mutual (relative) rotational deflections of such "rigid bodies" as molecules in a molecular crystal, or grains in a structured medium, usually modify the potential energy of the system and this potential energy function often exhibits the symmetries as described above. Formally the same problem may be investigated in $n$ dimensions, when $G={\rm SO}\left(n,\mathbb{R}\right)$ or its "quantum-spinorial" universal covering $G={\rm Spin}\left(n\right)$.

Incidentally, let us observe that the same properties are characteristic for the functions of the type:
\begin{equation}\label{eq.I.9}
f\left(g\right):={\rm Tr}\left(T\left(g^{k}\right)\right)= {\rm Tr}\left(T\left(g\right)^{k}\right),\qquad k\in\mathbb{Z}.
\end{equation}

If $G$ is non-Abelian, then, obviously, the mapping 
\begin{equation}
G\ni g\mapsto T\left(g^{k}\right)=T\left(g\right)^{k}\in{\rm GL}\left(X\right) 
\end{equation}
is not any longer the representation of $G$, nevertheless, the property (\ref{eq.I.8}) holds.

The situation becomes more complicated when $G={\rm GL}\left(3,\mathbb{R}\right)$ (generally ${\rm GL}\left(n,\mathbb{R}\right)$), i.e., when we consider the system of affinely-rigid, i.e., homogeneously deformable bodies. As mentioned, systems with such degrees of freedom may be good models of molecular crystals, granular media, and other physical objects. There are however some delicate problems concerning the possibility and physical status of dynamical affine invariance. This problem was discussed in many our former papers \cite{acta,all04,all05}. We discussed there a single affine body, therefore the study of affine invariance was restricted to the structure of kinetic energy. And even at this stage the problem of affine invariance was a delicate matter. Namely, according to the common belief, all non-relativistic physical theories contain the Euclidean spatial metric tensor $g$ as a kind of "controlling parameter". Specially-relativistic theories contain the pseudo-Euclidean metric of the normal-hyperbolic signature; formally its status is there exactly like in non-relativistic physics. The metric tensor is built up into the structure of kinetic energy and other physical quantities. In this way the assumed kinematical affine symmetry of degrees of freedom is dynamically broken and restricted to Euclidean subgroup (pseudo-Euclidean in specially-relativistic theories). But there is at least an academically-motivated interest in following the pattern of mechanics of rigid bodies and their systems. This motivates the search of models of kinetic energy invariant under the left/right action of the affine group. Such models exist \cite{acta,all04,all05} and are mathematically interesting. But of course the natural question appears as to their physical utility. In mentioned papers certain arguments were presented. The expected application fields usually have to do with situations in condensed matter, microstructure, and in all phenomena where the external influences become secondary in comparison with internal factors. Of course, one could try from the very beginning to think in the following way: The primary concept is that of kinetic energy or more generally of the kinetic term of Lagrangian. Metric-like tensors are byproducts of those kinetic terms, e.g., as configuration-dependent coefficients of quadratic forms of generalized velocities. Of course, one can reproach against such a reasoning on the basis of General Relativity, i.e., relativistic gravitation theory. This theory tells us that the metric tensor is a potential of the gravitational field and its special (pseudo-Euclidean, i.e., flat) form is simply something like the non-excited "ground state" of this field. So, one could argue, it is a primary physical factor and the structure of kinetic energies is its byproduct. But nevertheless one can admit possibility of situations where the internal factors are more essential, e.g., in defect theory. Moreover, one knows from the solid-state physics the situations where really the influence of "physical" metric on the inertial properties of bodies is at least remarkably reduced. For example, the effective inertia of electrons moving in crystals is characterized by the effective mass tensor \cite{Kit_71} which differs from the "physical" metric and is some byproduct of complicated interactions in the crystalline medium. On the opposite scale of physical phenomena, in astrophysics, one can also expect situations where the configuration and internal motion of highly condensed objects like, e.g., neutron stars is a dominant factor and perhaps in a good approximation to rigorous generally-relativistic considerations (based on Einstein equations) the surrounding metric field (which in turn influences the star dynamics) may be expected to be a simple algebraic byproduct of internal mechanical phenomena.

When systems of mutually interacting affine bodies are studied, some models of potential energy (usually binary ones) must be used, either somehow derived from the "micromodel" (usually in a difficult way) or phenomenologically postulated on the basis of some rough qualitative assumptions and symmetry principles. This motivates what is done below in the first section. Namely, the concept of relative (mutual) deformation tensors of pairs of affine bodies is discussed in details and the scalar invariants built of such tensors are constructed. They are expected to be basic scalar arguments of binary potentials of mutual interbody interactions. There is an essential novelty in comparison to deformation scalars of single affine bodies. Namely, there exist affinely-invariant scalars of mutual deformations. Because of this one can consider the hierarchy of interaction models according to their invariance groups, from Euclidean to affine ones. There is again an open question of applicability, nevertheless, the models are very interesting at least from the point of view of pure mathematics and rational mechanics. The analysis performed in the first section might seem too detailed. However, without a well-defined geometric and algebraic background, i.e., when relying merely on analytical matrix calculus, it is almost impossible to avoid misunderstandings and quite real mistakes.

\section{Linear Algebra of Mutual Deformation Tensors. Transformation Rules and Invariants}

We must begin with some algebraic and geometric formalism.

Let $U$, $V$ be finite-dimensional linear spaces and $U^{\ast}$, $V^{\ast}$ --- their duals (linear spaces of linear functions on $U$, $V$). We deal almost exclusively with real linear spaces, nevertheless, all the concepts quoted below are valid as well when $U$, $V$ are linear spaces over the complex field $\mathbb{C}$. In the complex case some additional concepts and structures appear (e.g., the complex-conjugate spaces, antidual spaces, etc.), however, we will not use them. The set of linear mappings from $U$ into $V$ will be denoted by ${\rm L}\left(U,V\right)$. Obviously, ${\rm L}\left(U,V\right)$ carries the canonical linear structure induced pointwisely from the target space $V$. The symbols $U^{\ast}$, $V^{\ast}$ are evident abbreviations for ${\rm L}(U,\mathbb{R})$, ${\rm L}(V,\mathbb{R})$ (${\rm L}(U,\mathbb{C})$ and ${\rm L}(V,\mathbb{C})$ in the complex case). The manifold of linear monomorphisms (injections) from $U$ to $V$ will be denoted by ${\rm LM}\left(U,V\right)$; obviously, it is an open submanifold of ${\rm L}\left(U,V\right)$ and is non-empty only if 
\begin{equation}
m:=\dim U\leq n:=\dim V. 
\end{equation}
The manifold of linear epimorphisms (projections) of $U$ onto $V$ will be denoted by ${\rm LE}\left(U,V\right)$. It is also an open submanifold of ${\rm L}\left(U,V\right)$ and is non-empty  only when $\dim U\geq \dim V$. The manifold of linear isomorphisms of $U$ onto $V$ will be denoted by ${\rm LI}\left(U,V\right)$. Obviously, 
\begin{equation}
{\rm LI}\left(U,V\right)={\rm LM}\left(U,V\right)={\rm LE}\left(U,V\right) 
\end{equation}
and it is non-empty only if 
\begin{equation}
\dim U=\dim V. 
\end{equation}
We shall use the obvious abbreviations ${\rm L}(U)$, ${\rm L}(V)$ for ${\rm L}(U,U)$, ${\rm L}(V,V)$ and the usual symbols of linear groups ${\rm GL}(U)$, ${\rm GL}(V)$ for ${\rm LM}(U,U)={\rm LE}(U,U)$, ${\rm LM}(V,V)={\rm LE}(V,V)$.

It is important for us that ${\rm L}(U)$, ${\rm L}(V)$ are associative algebras under the usual composition rule as a product and Lie algebras under the commutator operation:
\begin{equation}\label{eq.1}
[A,B]:=AB-BA.
\end{equation}
Similarly, ${\rm GL}(U)$, ${\rm GL}(V)$ are Lie groups and ${\rm L}(U)$, ${\rm L}(V)$ are their Lie algebras ${\rm GL}^{\prime}(U)$, ${\rm GL}^{\prime}(V)$ in a canonical way; the exponential mapping being understood literally as a power series in finite-dimensional associative algebras. ${\rm GL}^{+}(U)$, ${\rm GL}^{+}(V)$ denote the components of unity in ${\rm GL}(U)$, ${\rm GL}(V)$, i.e., the connected groups of linear transformations preserving any of two possible orientations in $U$, $V$. To be more in accord with popularly used terms, it is ${\rm GL}^{+}(U)$, ${\rm GL}^{+}(V)$ that are Lie groups and ${\rm L}(U)$, ${\rm L}(V)$ are their Lie algebras. The complements of ${\rm GL}^{+}(U)$, ${\rm GL}^{+}(V)$ to ${\rm GL}(U)$, ${\rm GL}(V)$ are denoted respectively by ${\rm GL}^{-}(U)$, ${\rm GL}^{-}(V)$. They consist of orientation-reversing mappings and are cosets of ${\rm GL}^{+}(U)$, ${\rm GL}^{+}(V)$. Obviously, determinants of ${\rm GL}^{+}$-, ${\rm GL}^{-}$-mappings are respectively positive and negative. Being finite-dimensional linear spaces (thus Abelian Lie groups in the additive sense), $U$ and $V$ are endowed with canonical translationally-invariant Lebesgue measures (special case of the Haar measure); they are unique up to a constant multiplier (normalization). The volume-preserving subgroups, i.e., unimodular groups ${\rm Um}(U)$, ${\rm Um}(V)$ consist of linear mappings the determinants of which have the absolute value equal to 1. Their special linear subgroups ${\rm SL}(U)$, ${\rm SL}(V)$ consist of orientation-preserving mappings, thus ones with determinants equal to one, 
\begin{equation}
{\rm SL}(U)={\rm Um}(U)\bigcap{\rm GL}^{+}(U) 
\end{equation}
and so for $V$. The subgroups ${\rm Um}(U)$, ${\rm Um}(V)$, ${\rm SL}(U)$, ${\rm SL}(V)$ are "structure-independent" in the sense that they do not depend on additional geometric objects in linear spaces. The same concerns dilatational groups 
\begin{eqnarray}
{\rm Dil}(U)=\{\lambda{\rm Id}_{U}:\lambda\neq 0\},&\quad& {\rm Dil}(V)=\{\lambda{\rm Id}_{V}:\lambda\neq 0\},\\ 
{\rm Dil}^{+}(U)=\{\lambda{\rm Id}_{U}:\lambda>0\},&\quad& {\rm Dil}^{+}(V)=\{\lambda{\rm Id}_{V}:\lambda>0\}. 
\end{eqnarray}
Let us also quote the orientation-reversing cosets: 
\begin{equation}
{\rm SL}^{-}(U)={\rm Um}(U)\bigcap{\rm GL}^{-}(U),\qquad {\rm Dil}^{-}(U)={\rm Dil}(U)\bigcap{\rm GL}^{-}(U) 
\end{equation}
(negative extension factors $\lambda$) and so for $V$. All other subgroups, at least practically useful ones, are defined as ones preserving some fixed geometric object or figure in an underlying linear space.

Below we will often deal with a pair of linear spaces $U$, $V$; their ordered bases (frames) will be denoted respectively by 
\begin{equation}
E=\left(\ldots,E_{A},\ldots\right),\qquad e=\left(\ldots,e_{i},\ldots\right). 
\end{equation}
The corresponding contravariant vector components are denoted as $u^{A}$, $v^{i}$,
\begin{equation}\label{eq.2}
u=u^{A}E_{A}\in U,\qquad v=v^{i}e_{i}\in V
\end{equation}
(summation convention applied).

The dual co-frames in $U^{\ast}$, $V^{\ast}$ will be denoted by 
\begin{equation}
E^{-1}=\left(\ldots,E^{A},\ldots\right),\qquad e^{-1}=\left(\ldots,e^{i},\ldots\right), 
\end{equation}
where obviously
\begin{equation}\label{eq.3}
E^{A}\left(E_{B}\right)=\langle E^{A},E_{B}\rangle=\delta^{A}{}_{B},\qquad e^{i}\left(e_{j}\right)=\langle e^{i},e_{j}\rangle=\delta^{i}{}_{j}.
\end{equation}
And, as usual, we use the lower-case convention for covariant vectors, i.e., linear functions:
\begin{equation}\label{eq.4}
f=f_{A}E^{A}\in U^{\ast},\qquad p=p_{i}e^{i}\in V^{\ast}.
\end{equation}
Therefore, as usual,
\begin{equation}\label{eq.5}
f(u)=\langle f,u \rangle=f_{A}u^{A,} \qquad p(v)=\langle p,v\rangle=p_{i}v^{i}.
\end{equation}

Matrix elements of linear mappings $\psi\in {\rm L}\left(U,V\right)$ are meant in the convention:
\begin{equation}\label{eq.6}
\psi E_{A}=e_{i}\psi^{i}{}_{A},
\end{equation}
therefore,
\begin{equation}\label{eq.7}
\psi\left(u^{A} E_{A}\right)=\left(\psi^{i}{}_{A}u^{A}\right)e_{i}, \qquad \psi(u)^{i}=\psi^{i}{}_{A}u^{A}.
\end{equation}
We shall often make use of canonical identifications:
\begin{equation}\label{eq.8}
{\rm L}\left(U,V\right)\simeq V\otimes U^{\ast}, \qquad U^{\ast\ast}\simeq U, \qquad {\rm etc.}
\end{equation}

Very often we must deal with various transposition concepts of linear mappings. This is the main motivation for this introductory review of symbols. Let us begin with the well-known concept of the conjugate mapping. Namely, with any $\psi\in  {\rm L}\left(U,V\right)$ there is canonically associated the mapping 
\begin{equation}
\psi^{\ast}\in {\rm L}(V^{\ast},U^{\ast})\simeq U^{\ast} \otimes V 
\end{equation}
given by
\begin{equation}\label{eq.9}
\psi^{\ast}p:=p\circ\psi \qquad {\rm for\ any} \qquad p\in V^{\ast}.
\end{equation}
In terms of coordinates:
\begin{equation}\label{eq.10}
\left(\psi^{\ast}p\right)_{A}=p_{i}\psi^{i}{}_{A}.
\end{equation}
Roughly speaking, the matrix of $\psi^{\ast}$ is transposed to that of $\psi$
in the usual analytical sense of matrix calculus. No additional geometric concept is  needed for defining the conjugation star-mapping (\ref{eq.9}), (\ref{eq.10}). And no wonder, the canonical isomorphism of linear spaces
\begin{equation}\label{eq.11}
{\rm L}\left(U,V\right)\simeq V\otimes U^{\ast}, \qquad {\rm L}(V^{\ast},U^{\ast})\simeq U^{\ast}\otimes V
\end{equation}
described by the star-operation (\ref{eq.9}), (\ref{eq.10}) is just the special case of  the natural isomorphism between $W\otimes Z$, $Z\otimes W$ ($W$, $Z$ being arbitrary linear spaces).

If $U$, $V$ have the same dimension and $\psi$ is an isomorphism, i.e., $\psi\in {\rm LI}\left(U,V\right)$, then we can define the contragradient mapping $\psi_{\ast}\in {\rm LI}(U^{\ast},V^{\ast})$ given by
\begin{equation}\label{eq.12}
\psi_{\ast}:=\left(\psi^{\ast}\right)^{-1}=\left(\psi^{-1}\right)^{\ast}.
\end{equation}
Analytically:
\begin{equation}\label{eq.13}
\left(\psi_{\ast}f\right)_{i}:=f_{A}\psi^{-1A}{}_{i},
\end{equation}
where obviously
\begin{equation}\label{eq.14}
\psi^{-1A}{}_{i}\psi^{i}{}_{B}=\delta^{A}{}_{B}, \qquad \psi^{i}{}_{A}\psi^{-1A}{}_{j}=\delta^{i}{}_{j}.
\end{equation}
For any $\psi\in {\rm L}\left(U,V\right)$, the upper-star-symbol $\psi^{\ast}$ is also used to denote the natural pull-back mapping of the covariant tensor algebra over $V$ onto that over $U$. Similarly, one uses $\psi_{\ast}$ to denote the natural extension (push-forward) of $\psi$ to the contravariant tensor algebra over $U$ (mapping it into the contravariant tensor algebra over $V$). If $U$, $V$ have the same dimension and $\psi$ is an isomorphism, then $\psi_{\ast}$, $\psi^{\ast}$ may be extended to the isomorphisms of the total tensor algebras, and obviously, $\psi_{\ast}=\left(\psi^{\ast}\right)^{-1}$ in this extended sense as well.

Obviously, for any  $\psi\in {\rm L}\left(U,V\right)$,  $\varphi\in {\rm L}(V,W)$, we have
\begin{equation}\label{eq.15}
(\varphi\psi)^{\ast}=\psi^{\ast}\varphi^{\ast},\qquad (\varphi\psi)_{\ast}=\varphi_{\ast}\psi_{\ast},
\end{equation}
i.e., respectively anti-representation and representation properties.

Below we shall often use the natural isomorphism of linear spaces ${\rm L}\left(U,V\right)^{\ast}$ and ${\rm L}(V,U).$ Namely, $\varphi\in {\rm L}(V,U)$ acts on $\psi\in {\rm L}\left(U,V\right)$ as a linear functional in the sense of the trace formula:
\begin{equation}\label{eq.16}
\langle \varphi, \psi\rangle:={\rm Tr}(\varphi\psi)={\rm Tr}(\psi \varphi)=\varphi^{i}{}_{A}\psi^{A}{}_{i}.
\end{equation}
In particular, ${\rm L}(U)^{\ast}$, ${\rm L}(V)^{\ast}$ are identical with ${\rm L}(U)$, ${\rm L}(V)$.

We were dealing as yet with amorphous linear spaces $U$, $V$ without any additional structure. From now on we assume $U$, $V$ to be endowed with some fixed metric tensors, i.e., symmetric non-degenerate bilinear forms,
\begin{equation}\label{eq.17}
g\in {\rm Sym}\left(V^{\ast}\otimes V^{\ast}\right)\subset V^{\ast}\otimes V^{\ast}, \qquad \eta\in {\rm Sym}\left(U^{\ast}\otimes U^{\ast}\right)\subset U^{\ast}\otimes U^{\ast}.
\end{equation}
They are analytically described by their components $g_{ij}$, $\eta_{AB}$ with respect to frames $e$, $E$:
\begin{equation}\label{eq.18}
\eta=\eta_{AB}E^{A}\otimes E^{B},\qquad g=g_{ij}e^{i}\otimes e^{j}.
\end{equation}
Their reciprocal contravariant tensors denoted by
\begin{equation}\label{eq.19}
g^{-1}\in {\rm Sym}\left(V\otimes V\right)\subset V\otimes V, \qquad \eta^{-1}\in {\rm Sym}\left(U\otimes U\right)\subset U\otimes U
\end{equation}
are analytically represented as
\begin{equation}\label{eq.20}
\eta=\eta^{AB}E_{A}\otimes E_{B},\qquad g=g^{ij}e_{i}\otimes e_{j},
\end{equation}
where, obviously,
\begin{equation}\label{eq.21}
\eta^{AC}\eta_{CB}=\delta^{A}{}_{B}, \qquad g^{ik}g_{kj}=\delta^{i}{}_{j}.
\end{equation}
In mechanical applications $\eta$, $g$ are positively definite, however,
at this stage the definiteness problem is not essential.

The pair of Euclidean spaces $\left(U,\eta\right)$, $\left(V,g\right)$ gives rise to a richer system of invariantly defined geometric objects. First of all, one can define new transposition concepts. They are metric-dependent, i.e., explicitly built of the metric tensors $\eta$, $g$. Just here they are different and more peculiar than the purely affine (linear) concept of the conjugate mapping $\psi^{\ast}$ defined above. Let us begin with coordinate-free definition and later
on we quote analytical formulas used in calculations.

The metric tensors $\eta$, $g$ give rise to the "index-lowering" operations producing covectors from vectors:
\begin{equation}\label{eq.22}
l_{\eta}\in {\rm L}\left(U,U^{\ast}\right),\qquad 
l_{g}\in {\rm L}\left(V,V^{\ast}\right).
\end{equation}
They are given by:
\begin{equation}\label{eq.23}
l_{\eta}u=\eta(u,\cdot)=\eta(\cdot,u)\in U^{\ast},\qquad 
l_{g}v=g(v,\cdot)=g(\cdot,v)\in V^{\ast}
\end{equation}
for any $u\in U$, $v\in V$. More explicitly,
\begin{equation}\label{eq.24}
\langle l_{\eta}u,w\rangle=\eta(u, w), \qquad \langle l_{g}v, z\rangle=g(v, z)
\end{equation}
for any $w\in U$, $z\in V$.

Due to non-singularity of $\eta$, $g$ (positive definiteness not yet necessary),
linear operations $l_{\eta}$, $l_{g}$ are invertible. Their inverses, i.e., "index-raising" operations will be denoted by
\begin{equation}\label{eq.25}
r_{\eta}=l_{\eta}^{-1}\in {\rm L}\left(U^{\ast},U\right),\qquad r_{g}=l_{g}^{-1}\in {\rm L}\left(V^{\ast},V\right).
\end{equation}

The $(\eta,g)$-transpose of $\psi\in {\rm L}\left(U,V\right)$ is defined as the linear mapping given by
\begin{equation}\label{eq.26}
{\rm L}(V,U)\ni\psi^{T(\eta,g)}:=
r_{\eta}\psi^{\ast}l_{g}=l_{\eta}{}^{-1}\psi^{\ast}l_{g}.
\end{equation}
If $(\eta,g)$ are obvious from the context, one writes for brevity $\psi^{T}$.
Analytically, $\psi^{T}$ is given by its matrix $\left[\psi^{TA}{}_{i}\right]$, where
\begin{equation}\label{eq.27}
\psi^{TA}{}_{i}=\eta^{AB}g_{ki}\psi^{k}{}_{B}=g_{ik}\psi^{k}{}_{B}\eta^{BA}.
\end{equation}

If the metrics $\eta$, $g$ are strictly Euclidean, i.e., positively definite, then there exist orthonormal frames, i.e., such ones that
\begin{equation}\label{eq.28}
\eta_{AB}=_{\ast}\eta^{AB}=_{\ast}\delta_{AB},\qquad 
g_{ij}=_{\ast}g^{ij}=_{\ast}\delta_{ij}.
\end{equation}
In such coordinates the matrices of $\psi^{\ast}$ and $\psi^{T}$ numerically coincide. It is no longer the case in general coordinates. And in the case of non-definite (pseudo-Euclidean) metrics such coordinates do not exist at all. Let us observe that literally speaking, $\psi^{\ast}$ and $\psi^{T}$ belong to different linear spaces, respectively 
\begin{equation}
{\rm L}\left(V^{\ast},U^{\ast}\right)\simeq U^{\ast}\otimes V,\qquad {\rm L}(V,U)\simeq U\otimes V^{\ast}. 
\end{equation}
They are mutually dual in a canonical way, however, they are not canonically isomorphic to each other. The isomorphism may be established only on the basis of metrics $\eta$, $g$.

The above constructions are valid for arbitrary dimensions of $U$, $V$; they need not be equal. If they are equal and if $\psi$ is an isomorphism, there exists the inverse mapping $\psi^{-1}\in {\rm L}(V,U)$ and the contragradient mapping 
\begin{equation}
\widetilde{\psi}=\psi^{\ast-1}=\psi^{-1\ast}\in {\rm L}\left(U^{\ast},V^{\ast}\right). 
\end{equation}
In certain formulas we shall need the $(\eta,g)$-transpose of $\psi^{-1}$, 
\begin{equation}\label{eq.29}
\left(\psi^{-1}\right)^{T(\eta,g)}\in{\rm L}\left(U,V\right)\simeq V\otimes U^{\ast}.
\end{equation}
It is given by
\begin{equation}\label{eq.30}
\left(\psi^{-1}\right)^{T}=r_{g}\psi^{-1\ast}l_{\eta}=
l_{g}{}^{-1}\psi^{-1\ast}l_{\eta},
\end{equation}
where the labels $\eta$, $g$ at the transposition symbol are omitted for brevity if there is no misunderstanding danger. The matrix of $\psi^{-1T}$ is given by
\begin{equation}\label{eq.31}
\left(\psi^{-1T}\right)^{i}{}_{A}=\eta_{AB}\psi^{-1B}{}_{j}g^{ji}.
\end{equation}

The above transposes depend explicitly on the both metric tensors $\eta\in U^{\ast}\otimes U^{\ast}$ and $g \in V^{\ast}\otimes V^{\ast}$. In certain formulas we use another type of transposes, depending only on one of the metrics.

So, for any 
\begin{equation}
\psi\in {\rm L}\left(U,V\right)\simeq V\otimes U^{\ast} 
\end{equation}
we define its $g$-transpose 
\begin{equation}
{}_{T(g)}\psi\in {\rm L}\left(V,U^{\ast}\right)\simeq U^{\ast} \otimes V^{\ast}; 
\end{equation}
again we omit the label $g$ when it is clear from the context. Then ${}_{T}\psi$ is given by
\begin{equation}\label{eq.32}
{}_{T}\psi:=\psi^{\ast}l_{g}
\end{equation}
and has the matrix with elements
\begin{equation}\label{eq.33}
{}_{T}\psi_{Ai}=g_{ij}\psi^{j}{}_{A}.
\end{equation}
It depends explicitly on $g$ but does not depend on $\eta$. If $g$ is positively definite (Euclidean) and the frame $e$ is orthonormal, then again the matrices of $\psi^{\ast}$ and ${}_{T}\psi$ are identical. Let us observe that obviously
\begin{equation}\label{eq.34}
\psi^{T}=r_{\eta}{}_{T}\psi=l_{\eta}^{-1}{}_{T}\psi
\end{equation}
and it is only here that the metric of $U$ appears.

If dimensions of $U$, $V$ are identical and $\psi\in{\rm L}(U, V)$ is an isomorphism, then for $\psi^{-1}\in {\rm L}(V,U)$ we define the transpose
\begin{equation}\label{eq.35}
{}_{T(\eta)}\psi^{-1}:=\psi^{-1\ast}l_{\eta}\in  {\rm L}\left(U,V^{\ast}\right)\simeq V^{\ast}\otimes U^{\ast}.
\end{equation}
Its matrix is given by
\begin{equation}\label{eq.36}
\left({}_{T}\psi^{-1}\right)_{iA}=\eta_{AB}\psi^{-1B}{}_{i}.
\end{equation}
Obviously,
\begin{equation}\label{eq.37}
\left(\psi^{-1}\right)^{T}=r_{g}{}_{T}\psi^{-1}.
\end{equation}
The transpose ${}_{T(\eta)}\psi^{-1}$ depends explicitly on $\eta$, but does not dependent on $g$.

It is seen that the purely analytical meaning of the transposed matrix may be misleading when one does not take care to the geometric status of objects represented by matrices. In various contexts all the mentioned transposes occur in mechanics of deformable bodies. Everything said above is applicable to the special case of automorphisms, when $U=V$ and $\eta=g$. Being mixed tensors, the elements of ${\rm L}(U)$, ${\rm L}(V)$ cannot be transposed in an affinely-invariant manner, in particular they cannot be either symmetric or antisymmetric. Their transpose is always an essentially metrical concept. When dealing with the general ${\rm L}\left(U,V\right)$, one is faced with this fact in an even more drastic way. With an obvious exception, of course, when $U=V^{\ast}$ or $U=V$. Then respectively
\begin{equation}\label{eq.38}
{\rm L}\left(V^{\ast},V\right)\simeq V\otimes V^{\ast\ast}\simeq V\otimes V, \qquad {\rm L}\left(V,V^{\ast}\right)\simeq V^{\ast}\otimes V^{\ast},
\end{equation}
and obviously the concepts of transposition, symmetry and skew-symmetry are well defined without any use of metrical concepts, thus, affinely-invariant.

Having fixed the metric tensors $\eta$, $g$ in $U$, $V$ we can distinguish within ${\rm LI}\left(U,V\right)$ the subset of isometries 
\begin{equation}
{\rm O}\left(U,\eta;V,g\right)\subset{\rm LI}\left(U,V\right), 
\end{equation}
consisting of isomorphisms preserving the metric structure, i.e., interrelating $\eta$, $g$:
\begin{equation}\label{eq.39}
\varphi\in {\rm O}\left(U,\eta;V,g\right):\qquad \eta=\varphi^{\ast}g,
\end{equation}
i.e., analytically
\begin{equation}\label{eq.40}
\eta_{AB}=g_{ij}\varphi^{i}{}_{A}\varphi^{j}{}_{B}.
\end{equation}

For general isomorphisms of $U$ onto $V$ the above relationship does not hold and the concept of deformation tensor appears in a natural way.

For any $\varphi\in{\rm LI}\left(U,V\right)$ the Green deformation tensor is defined in the usual way:
\begin{equation}\label{eq.41}
G[\varphi]=\varphi^{\ast}g\in {\rm Sym}\left(U^{\ast}\otimes U^{\ast}\right),
\end{equation}
i.e., analytically
\begin{equation}\label{eq.42}
G[\varphi]_{AB}=g_{ij}\varphi^{i}{}_{A}\varphi^{j}{}_{B}.
\end{equation}

Let us notice that $G[\varphi]$ does not involve the tensor $\eta$ at all, so strictly speaking, the term "deformation tensor" is not correctly used here, as no relationship between $\eta$ and $g$ occurs here at all. It would be more adequate to call $G[\varphi]$ the "metric" of $U$ which is $\varphi$-induced from $\eta$. But for traditional reasons we follow the terms used in the Eringen book \cite{Erin_b1,Erin_b2} (although, to be honest, an "infinity" of other names occurs also in the literature; we stick to the Eringen conventions). Using the above transposition concepts we can simply write
\begin{equation}\label{eq.43}
G[\varphi]={}_{T}\varphi\varphi\in {\rm L}\left(U,U^{\ast}\right)\simeq U^{\ast}\otimes U^{\ast};
\end{equation}
as mentioned, to reduce the crowd of symbols, we refrain from using the unambiguous expression ${}_{T(g)}\varphi\varphi$, because $g$ is assumed to be known and fixed.

For certain purposes we use also the mixed tensor
\begin{equation}\label{eq.45}
\widehat{G}[\varphi]=\varphi^{T}\varphi\in {\rm L}(U)\simeq U\otimes U^{\ast},
\end{equation}
analytically
\begin{equation}\label{eq.46}
\widehat{G}[\varphi]^{A}{}_{B}=\eta^{AC}G[\varphi]_{CB}=
\eta^{AC}g_{ij}\varphi^{i}{}_{C}\varphi^{j}{}_{B}.
\end{equation}
Also here we do not use the more correct symbol
\begin{equation}\label{eq.47}
\widehat{G}[\varphi]=\varphi^{T(\eta,g)}\varphi
\end{equation}
to avoid the "multi-floor" crowd of characters. If one uses rectilinear orthonormal coordinates in $U$, $V$, then the symbol $\varphi^{T}\varphi$ becomes exactly the standard notation used in elasticity.

Some warning: one uses also the quantity
\begin{equation}\label{eq.48}
\widetilde{G}[\varphi]\in U \otimes U
\end{equation}
given by the $\eta$-raising of indices, analytically
\begin{equation}\label{eq.49}
\widetilde{G}[\varphi]^{AB}=\eta^{AC}\eta^{BD}G[\varphi]_{CD}.
\end{equation}
It must be not confused with the contravariant reciprocal,
\begin{equation}\label{eq.50}
G[\varphi]^{-1}\in U \otimes U,
\end{equation}
where
\begin{equation}\label{eq.51}
\left(G[\varphi]^{-1}\right)^{AC}G[\varphi]_{CB}=\delta^{A}{}_{B}.
\end{equation}
Both of them are used and coincide only when $\varphi$ is an isometry, i.e., when $\varphi\in{\rm O}\left(U,\eta;V,g\right)$, reducing then to $\eta^{-1}$, the contravariant reciprocal of $\eta$. And of course, in this case, when there is no deformation, $G[\varphi]$ itself becomes $\eta$. Then it is convenient to use the deformation measure which vanishes. It is the Lagrange deformation tensor $E[\varphi]\in U^{\ast}\otimes U^{\ast}$:
\begin{equation}\label{eq.52}
E[\varphi]:=\frac{1}{2}\left(G[\varphi]-\eta\right).
\end{equation}

Unlike $G[\varphi]$, ${\rm E}(\varphi)$ depends explicitly on both $g$ and $\eta$, so in this case the term "deformation tensor" is correctly and literally used. Just as previously, the objects with $\eta$-raised indices are used, 
\begin{equation}
\widehat{E}[\varphi]\in U\otimes U^{\ast},\qquad \widetilde{E}[\varphi]\in U \otimes U; 
\end{equation}
analytically
\begin{equation}\label{eq.53}
\widehat{E}[\varphi]^{A}{}_{B}=\eta^{AC}E[\varphi]_{CB},\qquad \widetilde{E}[\varphi]^{AB}=\eta^{AC}\eta^{BD}E[\varphi]_{CD}.
\end{equation}
Obviously, in general something like $E[\varphi]^{-1}$, $\widehat{E}[\varphi]^{-1}$ need not be well defined and even if it accidentally happens to exist, it is rather non-useful to anything. One can suspect that perhaps some profit might follow from using the quantity 
\begin{equation}
\mathcal{E}[\varphi]\in U \otimes U^{\ast}\simeq {\rm L}(U) 
\end{equation}
given implicitly by 
\begin{equation}\label{eq.54}
\widehat{G}[\varphi]:=\exp\left(2\mathcal{E}[\varphi]\right).
\end{equation}
Obviously, for "almost isometric" $\varphi$ ("small deformations") the quantities $\mathcal{E}[\varphi]$, $\widehat{E}[\varphi]$ asymptotically coincide ($\exp(x)\approx 1+x$ for $x\approx 0$).

Let us observe that $G[\varphi]$, $\widehat{G}[\varphi]$, $\widetilde{G}[\varphi]$, $E[\varphi]$, $\widehat{E}[\varphi]$, $\widetilde{E}[\varphi]$ exist also when $\varphi$ is not an isomorphism, in particular, when $U$, $V$ have different dimensions \cite{Roz_05}.

For any $\psi\in{\rm LI}\left(U,V\right)$ the covariant Cauchy deformation tensor $C[\psi]$ is defined as follows:
\begin{equation}\label{eq.55}
C[\psi]=\psi_{\ast}\eta =\psi^{-1\ast}\eta\in {\rm Sym}\left(V^{\ast}\otimes V^{\ast}\right).
\end{equation}
Analytically
\begin{equation}\label{eq.56}
C[\psi]_{ij}=\eta_{AB}\psi^{-1A}{}_{i}\psi^{-1B}{}_{j}.
\end{equation}
Just the same remark as previously: it does not depend on $g$, does not relate $\eta$ to $g$, so it is not a "true" deformation tensor. It is a "metric-like" tensor in $V$ which is $\psi$-induced by $\eta$. Using the transposition symbol we can write that
\begin{equation}\label{eq.57}
C[\psi]={}_{T}\psi^{-1}\psi^{-1},
\end{equation}
where again for brevity we write simply $T$ instead of $T(\eta)$.

Again some byproducts of $C[\psi]$ are important and widely used. Let us begin with the contravariant inverse $C[\psi]^{-1}\in {\rm Sym}\left(V\otimes V\right)$, also $g$-independent one,
\begin{equation}\label{eq.58}
C[\psi]^{-1}=\psi_{\ast}\eta^{-1},
\end{equation}
i.e., analytically:
\begin{equation}\label{eq.59}
C[\psi]^{-1ij}=\psi^{i}{}_{A}\psi^{j}{}_{B}\eta^{AB}.
\end{equation}
The $g$-dependent byproducts of $C[\psi]$ denoted by
$\widehat{C}[\psi]\in {\rm L}(V)\simeq V\otimes V^{\ast}$ and $\widetilde{C}[\psi] \in V \otimes V$ are analytically given by
\begin{equation}\label{eq.60-61}
\widehat{C}[\psi]^{i}{}_{j}=g^{ik}C[\psi]_{kj}=
g^{ik}\eta_{AB}\psi^{-1A}{}_{k}\psi^{-1B}{}_{j},\quad
\widetilde{C}[\psi]^{ij}=g^{ik}g^{jl}C[\psi]_{kl}.
\end{equation}
Obviously,
\begin{equation}\label{eq.62}
\widehat{C}[\psi]=\psi^{-1T}\psi^{-1},\qquad \widehat{C}[\psi]^{-1}=\psi\psi^{T}.
\end{equation}
When using orthonormal coordinates and matrix language, the last expression coincides with the known analytic formula from elasticity textbooks. Just like the following one:
\begin{equation}\label{eq.63}
C[\psi]^{-1}=\psi\left({}_{T}\psi\right).
\end{equation}
Let us notice some important and delicate point. The covariant (thus, metric-like) Cauchy tensor $C[\psi]$ does exist only when $\psi$ is an isomorphism (thus, in particular, $U$, $V$ have the same dimension). But the contravariant inverse is built algebraically of $\psi$ alone, without any reliance on $\psi^{-1}$. So, strictly speaking, the primary quantity is just the contravariant push-forward of $\eta^{-1}$:
\begin{equation}\label{eq.64}
\mathcal{J}[\psi]:=\psi_{\ast}\eta^{-1},\qquad \mathcal{J}[\psi]^{ij}=\psi^{i}{}_{A}\psi^{j}{}_{B}\eta^{AB},
\end{equation}
i.e., using the transpose symbol again
\begin{equation}\label{eq.65}
\mathcal{J}[\psi]=\psi\left({}_{T}\psi\right).
\end{equation}
If $U$, $V$ have the same dimension and $\psi$ is an isomorphism, then we can obtain the covariant Cauchy tensor as
\begin{equation}\label{eq.66}
C[\psi]=\mathcal{J}[\psi]^{-1}.
\end{equation}
For the general, not necessarily isomorphic $\psi$, the primary quantity is $\mathcal{J}[\psi]$ and it need not be reciprocal to anything (being in general singular).

If $\psi$ is an isometry, then $C[\psi]=g$. Just as previously, it is often convenient to use the deformation measure vanishing for isometries. The Euler deformation tensor $e[\psi]\in{\rm Sym}\left(V^{\ast}\otimes V^{\ast}\right)$ is defined as follows:
\begin{equation}\label{eq.67}
e[\psi]=\frac{1}{2}\left(g-C[\psi]\right).
\end{equation}
Notice the seeming difference in sign convention in comparison with $E[\psi]$. It is not accidental, the both conventions are compatible with each other on the level of so-called infinitesimal deformation tensors. It would be, maybe, convenient to use the quantity 
\begin{equation}
\varepsilon[\psi]\in {\rm L}(V)\simeq V\otimes V^{\ast} 
\end{equation}
given implicitly by
\begin{equation}\label{eq.68}
\widehat{C}[\psi]=\exp\left(-2\varepsilon[\psi]\right).
\end{equation}

These were deformation tensors of linear mappings acting from $U$ to $V$; they were always related to some fixed metric structures in $U$, $V$. The next important concept is one of the mutual deformation of two linear mappings from $U$ to $V$.

Let us begin from some digression. In affine spaces with every pair of points there is associated a unique translation vector relating them to each other. Similarly, with every ordered pair of group elements there are associated two ones which acting as left or right translations interrelate them.

Let $\varphi,\psi\in{\rm LI}\left(U,V\right)$ be two arbitrary isomorphisms. Following the mentioned concept of translation vector and left or right translations in group manifolds we introduce the following concepts of mutual displacements in ${\rm LI}\left(U,V\right)$:
\begin{equation}\label{eq.69}
\Gamma[\psi,\varphi]=\psi^{-1}\varphi\in{\rm GL}(U),\qquad \Sigma[\psi,\varphi]=\varphi\psi^{-1}\in{\rm GL}(V).
\end{equation}
Obviously,
\begin{equation}
\Gamma[\psi,\varphi]=\Gamma[\varphi,\psi]^{-1},\qquad \Sigma[\psi,\varphi]=\Sigma[\varphi,\psi]^{-1},
\end{equation}
i.e., these operations are "skew-symmetric" in the group sense with respect to their arguments, just like the translation vector in affine spaces. Obviously, the objects $\Gamma$, $\Sigma$ are well define only for isomorphisms $\psi$, $\varphi$. And they are equal to the group identity if and only if $\psi=\varphi$. Incidentally, it may be also convenient to use "infinitesimal" versions of displacement objects, $\gamma$, $\sigma$, where 
\begin{equation}\label{eq.70}
\Gamma[\psi,\varphi]={\rm Id}_{U}+\gamma,\qquad 
\Sigma[\psi,\varphi]={\rm Id}_{V}+\sigma,
\end{equation}
vanishing when $\psi=\varphi$. And again perhaps more convincing geometrically
would be "logarithms" of $\Gamma$, $\Sigma$, i.e., $\alpha$, $\beta$ defined by
\begin{equation}\label{eq.71}
\Gamma[\psi,\varphi]=\exp\left(\alpha[\psi,\varphi]\right),\qquad \Sigma[\psi,\varphi]=\exp\left(\beta[\psi,\varphi]\right).
\end{equation}
As we remember, the Green deformation tensor $G[\psi]$ and contravariant Cauchy tensor $\mathcal{J}[\psi]$ do exist for the general $\psi$, not necessarily isomorphic ones. It is natural to ask for their analogues for pairs of mappings $\psi,\varphi\in{\rm LI}\left(U,V\right)$. 

The two-argument analogue of the Green tensor is given by
\begin{equation}\label{eq.72}
G[\psi,\varphi]\in U^{\ast}\otimes U^{\ast}\simeq 
{\rm L}\left(U,U^{\ast}\right),
\end{equation}
where
\begin{equation}\label{eq.73}
G[\psi,\varphi]:={}_{T}\psi\varphi
\end{equation}
with the above sense of symbols. Analytically
\begin{equation}\label{eq.74}
G[\psi,\varphi]_{AB}=g_{ij}\psi^{i}{}_{A}\varphi^{j}{}_{B}.
\end{equation}
Obviously, the symmetry is now replaced by
\begin{equation}\label{eq.75}
G[\psi,\varphi]^{T}=G[\varphi,\psi],
\end{equation}
here with the ${\rm GL}(U)$-invariant sense of transposition (twice covariant tensor). Just like $G[\psi]$, $G[\psi,\varphi]$ depends on $g$ but not on $\eta$. The obvious $\eta$-dependent object is given by
\begin{equation}\label{eq.76}
\widehat{G}[\psi,\varphi]:=\psi^{T}\varphi\in {\rm L}(U)\simeq U\otimes U^{\ast},
\end{equation}
i.e., analytically
\begin{equation}\label{eq.77}
\widehat{G}[\psi, \varphi]^{A}{}_{B}=\eta^{AC}G[\psi,\varphi]_{CB}=
\eta^{AC}g_{ij}\psi^{i}{}_{C}\varphi^{j}{}_{B}.
\end{equation}
Again in the metrically-transposed sense we have that
\begin{equation}\label{eq.78}
\widehat{G}[\varphi,\psi]=\widehat{G}[\psi,\varphi]^{T}.
\end{equation}
The mutual Cauchy tensor
\begin{equation}\label{eq.79}
C[\psi,\varphi]\in V^{\ast}\otimes V^{\ast}\simeq {\rm L}\left(V,V^{\ast}\right)
\end{equation}
does not depend on $g$ but is explicitly $\eta$-dependent,
\begin{equation}\label{eq.80}
C[\psi,\varphi]:={}_{T}\varphi^{-1}\psi^{-1},
\end{equation}
i.e., analytically
\begin{equation}\label{eq.81}
C[\psi,\varphi]_{ij}=\eta_{AB}\varphi^{-1A}{}_{i}\psi^{-1B}{}_{j}.
\end{equation}

As previously, the transposition of arguments is identical with the affinely-invariant tensor transposition
\begin{equation}\label{eq.82}
C[\psi,\varphi]^{T}=C[\varphi,\psi].
\end{equation}
The mixed-tensor representation
\begin{equation}\label{eq.83}
\widehat{C}[\psi,\varphi]\in V \otimes V^{\ast}\simeq {\rm L}(V),
\end{equation}
or analytically
\begin{equation}\label{eq.84}
\widehat{C}[\psi,\varphi]^{i}{}_{j}=g^{ik}C[\psi,\varphi]_{kj}=
g^{ik}\eta_{AB}\varphi^{-1A}{}_{k}\psi^{-1B}{}_{j},
\end{equation}
depends explicitly on both $\eta$ and $g$, whereas the permutation of its arguments results in the $(\eta,g)$-metrical transposition. The inverse contravariant of $C[\psi,\varphi]$
\begin{equation}\label{eq.85}
C[\psi,\varphi]^{-1}=\psi{}_{T}\varphi\in V\otimes V\simeq {\rm L}\left(V^{\ast},V\right)
\end{equation}
is analytically given by
\begin{equation}\label{eq.86}
C[\psi,\varphi]^{-1ij}=\psi^{i}{}_{A}\varphi^{j}{}_{B}\eta^{AB}.
\end{equation}
Again we see that now this contravariant tensor is again more natural then the covariant one $C[\psi,\varphi]$, because no inverses of $\psi$, $\varphi$ are used. Therefore, we can use $\mathcal{J}[\psi,\varphi]$, the analogue of the one-argument $\mathcal{J}[\psi]$. Even if $\psi$, $\varphi$ are not isomorphic, (even if they cannot be so because of different dimensions of $U$, $V$), we can define
\begin{equation}\label{eq.87}
\mathcal{J}[\psi,\varphi]\in V\otimes V\simeq {\rm L}\left(V^{\ast},V\right)
\end{equation}
just as 
\begin{equation}\label{eq.88}
\mathcal{J}[\psi,\varphi]^{ij}=\psi^{i}{}_{A}\varphi^{j}{}_{B}\eta^{AB}.
\end{equation}
This object is primary and if $\psi$, $\varphi$ are isomorphisms, then
\begin{equation}\label{eq.89}
\mathcal{J}[\psi,\varphi]=C[\psi,\varphi]^{-1}.
\end{equation}
Unlike this, $G[\psi,\varphi]$ is natural and its contravariant inverse
$G[\psi,\varphi]^{-1}$ does exist only if $\psi$, $\varphi$ are isomorphisms.

In analogy to (\ref{eq.45}), (\ref{eq.49}), (\ref{eq.60-61}) one uses also the tensors 
\begin{equation}
\widetilde{G}[\psi,\varphi],\qquad \widehat{G}[\psi,\varphi],\qquad \widetilde{C}[\psi,\varphi],\qquad \widehat{C}[\psi,\varphi].
\end{equation}

Let us observe that if $\psi$ is an isometry, $\psi\in{\rm O}\left(U,\eta;V,g\right)$, then 
\begin{equation}
\widehat{G}[\psi,\varphi]=\Gamma[\psi,\varphi]. 
\end{equation}
Similarly, if $\varphi$ is an isometry, then 
\begin{equation}
\widehat{C}[\psi,\varphi]=\Sigma[\psi,\varphi]. 
\end{equation}
It is rather intuitive that in those special cases the mutual (two-argument) deformation tensors coincide with the group-theoretic "displacements" $\Gamma$, $\Sigma$.

The next important problem is one concerning transformation properties of the above objects. The associative algebras ${\rm L}(U)$, ${\rm L}(V)$ may be considered as semigroups acting on ${\rm L}\left(U,V\right)$ according to the rules:
\begin{eqnarray}
{\rm L}\left(U,V\right)\ni\varphi &\mapsto& A\varphi,\qquad A\in{\rm L}(V),\label{eq.90a}\\
{\rm L}\left(U,V\right)\ni\varphi &\mapsto& \varphi B,\qquad B\in{\rm L}(U).\label{eq.90b}
\end{eqnarray}

When restrict ourselves to $A$, $B$ running over the groups ${\rm GL}(V)$, ${\rm GL}(U)$, then obviously ${\rm LI}\left(U,V\right)$ is preserved under mappings (\ref{eq.90a}), (\ref{eq.90b}) which form then the transformation groups acting transitively, freely and effectively on ${\rm LI}\left(U,V\right)$. Let us observe however that if dimensions of $U$, $V$ are not equal, then ${\rm GL}(V)$ acts also transitively on the manifold of monomorphisms ${\rm LM}\left(U,V\right)$, but the action of ${\rm GL}(U)$ is non-transitive \cite{Roz_05,Roz_10}. 
One should also remember that the action of ${\rm GL}(V)\times{\rm GL}(U)$ on ${\rm LM}\left(U,V\right)$ given by
\begin{equation}\label{eq.91}
{\rm LM}\left(U,V\right)\ni\varphi \mapsto A\varphi B,\qquad A\in{\rm GL(V)},\qquad B\in{\rm GL}(U)
\end{equation}
is non-effective. For example, if $\dim U=\dim V$, the kernel of non-effectivity is given by the following subgroup:
\begin{equation}\label{eq.92}
\left\{\left(\lambda{\rm Id}_{V},\lambda^{-1}{\rm Id}_{U}\right):\lambda\in\mathbb{R}\backslash\{0\}\right\}\subset
{\rm GL}(V)\times{\rm GL}(U).
\end{equation}
The action of subgroups of ${\rm GL}(V)$, ${\rm GL}(U)$ on ${\rm L}(U, V)$ is also interesting in various problems. Obviously, the restriction of (\ref{eq.91}) to the subgroup ${\rm SL}(V)\times{\rm SL}(U)$ is an effective action if both $V$, $U$ are odd-dimensional. If they are even-dimensional, the kernel of non-effectivity is the two-element subgroup of (\ref{eq.92}) corresponding to $\lambda=\pm 1$. If (\ref{eq.91}) is restricted to the group
${\rm UL}(V)\times{\rm UL}(U)$, where ${\rm UL}$ are unimodular subgroups of ${\rm GL}$ (determinants with the modulus equal one), then, obviously, the same two-element non-effectivity does exist independently of dimensions of $U$, $V$.

Let us remind that the Lie algebras  ${\rm SL}(V)^{\prime}$, ${\rm SL}(U)^{\prime}$ consist of traceless linear operators.

In many applications, including elasticity and other branches of continuum mechanics, we are particularly interested in the restriction of (\ref{eq.91}) to the isometry groups products ${\rm O}\left(V,g\right)\times{\rm O}\left(U,\eta\right)$, where obviously ${\rm O}\left(V,g\right)$ is an abbreviation for ${\rm O}(V,g;V,g)$; the same concerns ${\rm O}(U,\eta)$. These are subgroups of ${\rm U}\left(V,g\right)$, ${\rm U}\left(U,\eta\right)$, so in the case of even dimension their action through (\ref{eq.91}) has also the two-element non-effectiveness. Replacing the full orthogonal groups by their proper (determinant-one) subgroups 
\begin{equation}
{\rm SO}\left(V,g\right)={\rm O}\left(V,g\right)\bigcap {\rm SL}(V),\qquad {\rm SO}\left(U,\eta\right)={\rm O}\left(U,\eta\right)\bigcap {\rm SL}(U), 
\end{equation}
one obtains again the effective action in (\ref{eq.91}). Let us remind that the Lie algebras ${\rm SO}\left(V,g\right)^{\prime}$, ${\rm SO}\left(U,\eta\right)^{\prime}$ of the proper orthogonal groups consist respectively of the $g$- and $\eta$-skew-symmetric operators, i.e.,
\begin{equation}\label{eq.93}
g\left(Ax,y\right)=-g\left(x,Ay\right),\qquad \eta\left(Bu,v\right)=-\eta\left(u,Bv\right),
\end{equation} 
analytically
\begin{equation}\label{eq.94}
A^{i}{}_{j}=-A_{j}{}^{i}:=-g^{ik}g_{jl}A^{l}{}_{k},\qquad B^{K}{}_{L}=-B_{L}{}^{K}:=-\eta^{KM}\eta_{LN}B^{N}{}_{M}.
\end{equation}
Let us now discuss the transformation properties of the deformation and bi-deformation tensors. This will be necessary in symmetry studies of elasticity and other branches of continuum mechanics.

For any linear isometry $A\in{\rm O}\left(V,g\right)$ we have obviously 
\begin{equation}\label{eq.95}
G\left[A\psi,A\varphi\right]=G[\psi,\varphi],\qquad G\left[A\psi\right]=G[\psi];
\end{equation}
the latter is just the special case of the first one, because obviously 
\begin{equation}
G[\psi,\psi]=G[\psi].
\end{equation}

And for an arbitrary $A\in{\rm GL}(V)$ there is no concise expression for $G\left[A\psi,A\varphi\right]$ in terms of $G[\psi,\varphi]$ (also if $\psi=\varphi$). But for any $B\in{\rm GL}(U)$ the following nice formula holds:
\begin{equation}\label{eq.96}
G\left[\psi B,\varphi B\right]=B^{\ast}G[\psi,\varphi], 
\end{equation}
i.e., analytically
\begin{equation}\label{eq.97}
G\left[\psi B,\varphi B\right]_{KL}=G[\psi,\varphi]_{MN}B^{M}{}_{K}B^{N}{}_{L}. 
\end{equation}
For the Cauchy-type tensors we have so-to-speak the dual transformation rules. 
So, for any isometry $B\in{\rm O}\left(U,\eta\right)$
\begin{equation}\label{eq.98}
C\left[\psi B,\varphi B\right]=C[\psi,\varphi],\qquad {\rm i.e.,}\qquad 
\mathcal{J}\left[\psi B,\varphi B\right]=\mathcal{J}[\psi,\varphi],
\end{equation}
and for a general $B\in{\rm  GL}(U)$ there is no explicit expression for $C\left[\psi B, \varphi B\right]$ ($\mathcal{J}\left[\psi B,\varphi B\right]$) as an algebraic function of $C[\psi,\varphi]$ $(\mathcal{J}[\psi,\varphi])$.
But for any $A\in{\rm GL}(V)$ the following analogue of (\ref{eq.96}) holds:
\begin{equation}\label{eq.99}
C\left[A\psi,A\varphi\right]=A_{\ast}C[\psi,\varphi],\qquad {\rm i.e.,}\qquad \mathcal{J}\left[A\psi,A\varphi\right]=A_{\ast}\mathcal{J}[\psi,\varphi].
\end{equation}
Analytically 
\begin{equation}
C\left[A\psi,A\varphi\right]_{ij}=
C\left[\psi,\varphi\right]_{kl}A^{-1k}{}_{i}A^{-1l}{}_{j},\label{eq.100}
\end{equation}
i.e., 
\begin{equation}
\mathcal{J}\left[A\psi,A\varphi\right]^{ij}=
A^{i}{}_{k}A^{j}{}_{l}\mathcal{J}\left[\psi,\varphi\right]^{kl}.\label{eq.101}
\end{equation}
And finally, it is clear that the non-metrical mutual displacements
$\Gamma\left[\psi,\varphi\right]$, $\Sigma\left[\psi,\varphi\right]$
transform in the following way: 

For any $A\in{\rm GL}(V)$ 
\begin{equation}
\Gamma\left[A\psi,A\varphi\right]=\Gamma\left[\psi,\varphi\right],\qquad
\Sigma\left[A\psi,A\varphi\right]=A\Sigma\left[\psi,\varphi\right]A^{-1}.
\label{eq.102}
\end{equation}
and for any $B\in{\rm GL}(U)$ 
\begin{equation}
\Gamma\left[\psi B,\varphi B\right]=B^{-1}\Gamma\left[\psi,\varphi\right]B,\qquad
\Sigma\left[\psi B,\varphi B\right]=\Sigma\left[\psi,\varphi\right].\label{eq.103}
\end{equation}
These are invariance and the adjoint/co-adjoint rules. 

In all physical applications, including ones in mechanics of deformable
bodies, one needs scalar quantities built of tensors; their "invariants"
roughly speaking. Let us quote the basic ones built of the tensors
$G$, $C$, $\Gamma$, $\Sigma$. The standard (as a matter of fact canonical
and unique) way of constructing scalars from mixed second-order tensors
is to take the trace of their powers (as linear mappings). According
to the Cayley-Hamilton theorem it is sufficient to take $n$ of them
if the dimension of space equals $n$; any other will be some function
of those ones. Obviously, there is no way to construct well-defined
scalars from a single twice-covariant or twice-contravariant tensor (only scalar densities may be built of them). What is then commonly called the invariant
is a scalar quantity built of the pair of tensors, one of them being
implicitly assumed. This fact is often overlooked when one works analytically
in $\mathbb{R}^{n}$ using formally the matrix language. 

Let us begin with the basic affine invariants
\begin{equation}
\mathcal{M}_{a}\left[\psi,\varphi\right]=
\textrm{Tr}\left(\Gamma\left[\psi,\varphi\right]^{a}\right)=
\textrm{Tr}\left(\Sigma\left[\psi,\varphi\right]^{a}\right),\qquad a=1,\ldots,n,\label{eq.104}
\end{equation}
where, obviously, $X^{a}$ denotes $n$-th algebraic power of the linear mapping $X$. These invariants are really affine in the sense that 
\begin{equation}
\mathcal{M}_{a}\left[A\psi B,A\varphi B\right]=\mathcal{M}_{a}\left[\psi,\varphi\right],\qquad a=1,\ldots,n,\label{eq.105}
\end{equation}
for any $A\in{\rm GL}(V),$ $B\in{\rm GL}(U)$, i.e., they are invariant under
the total group (\ref{eq.91}) and do not assume any metrical or other additional
structures in the linear spaces $V$, $U$. Let us observe that these quantities
are essentially two-argument ones and, just like $\Gamma$, $\Sigma$
themselves, they trivialize (become constant) when $\psi=\varphi$.
There exist also scalars which do not degenerate in this sense, however,
for the price that the above affine invariance is lost. They preassume
the metric tensors $g\in V^{\ast}\otimes V^{\ast}$, $\eta\in U^{\ast}\otimes U^{\ast}$ and depend explicitly of them. They are given by
\begin{equation}
\mathcal{K}_{a}\left[\psi,\varphi\right]:=
{\rm Tr}\left(\widehat{G}\left[\psi,\varphi\right]^{a}\right),\qquad a=1,\ldots,n.\label{eq.106}
\end{equation}
One can easily check that 
\begin{equation}
\mathcal{K}_{a}\left[\psi,\varphi\right]=
{\rm Tr}\left(C\left[\psi,\varphi\right]^{-a}\right)=
{\rm Tr}\left(\mathcal{J}\left[\psi,\varphi\right]^{a}\right).\label{eq.107}
\end{equation}
If $\psi=\varphi$, the resulting quantities 
\begin{equation}
\mathcal{K}_{a}\left[\psi\right]:=
\mathcal{K}_{a}\left[\psi,\psi\right]\label{eq.108}
\end{equation}
become that is called in elasticity deformation invariants. It is
clear that 
\begin{equation}
\mathcal{K}_{a}\left[A\psi B,A\varphi B\right]=
\mathcal{K}_{a}\left[\psi,\varphi\right],\qquad a=1,\ldots,n,\label{eq.109}
\end{equation}
for any $A\in{\rm O}\left(V,g\right)$, $B\in{\rm O}\left(U,\eta\right)$, i.e., they are invariant
under (\ref{eq.91}) restricted to ${\rm O}\left(V,g\right)\times{\rm O}\left(U,\eta\right)$. Any other
scalar built of $\left(\psi,\varphi\right)$ and showing this rotational
invariance is a function of $\mathcal{K}_{a}\left[\psi,\varphi\right]$,
so they form a functional basis of the space of rotational invariants.
In mechanics of continua one uses also various other choices of basic
invariants.

{\bf Remark:} More precisely, $\mathcal{K}_{a}$ span functionally
the space of scalars invariant under ${\rm O}\left(V,g\right)\times{\rm O}\left(U,\eta\right)$ but
not under any larger subgroup of (\ref{eq.91}); let us observe that $\mathcal{M}_{a}\left[\psi,\varphi\right]$ being ${\rm GL}(V)\times{\rm GL}(U)$-invariant are automatically invariant under also under the subgroup ${\rm O}\left(V,g\right)\times{\rm O}\left(U,\eta\right)$, but are not functions of $\mathcal{K}_{a}\left[\psi,\varphi\right]$.

The usual deformation invariants $\mathcal{K}_{a}\left[\psi\right]$
(or any other system of $n$ functionally independent expressions
built of them) tell us, roughly speaking, how strongly the elastic
body is strained, what are the scalar magnitudes of deformation, i.e.,
stretchings, but tell us nothing about how is this deformation
oriented in space (compare: velocity vector and its absolute value).
The same concerns the two-argument quantities $\mathcal{K}_{a}\left[\psi,\varphi\right]$ characterizing the scalar multi-magnitudes of mutual "displacements" between $\psi$ and $\varphi$.

It is worth to mention about the volume orientation problems. Being
finite-dimensional linear spaces (thus, Abelian Lie groups under the
addition of vectors), $U$, $V$ have canonical-up-to-normalization
translationally-invariant measures, namely the Lebesgue measures (the
special case of Haar measures). This has nothing to do with other
metric-type concepts like distances and angles. However, when some
metrics $\eta\in U^{\ast}\otimes U^{\ast}$, $g\in V^{\ast}\otimes V^{\ast}$ 
are fixed, then there is a canonical normalization of those measures. The orientation problem is more complicated. Any linear space has two possible orientations and fixing $\eta$, $g$ does not distinguish any of them. Let us fix
some orientations $\omega$, $\ell$ respectively in $U$, $V$. They
are equivalence classes of $n$-forms, i.e., elements of $\Lambda^{n}U^{\ast}$
$\Lambda^{n}V^{\ast}$; the equivalence given by multiplication by positive
scalars. The manifold of isomorphisms ${\rm LI}\left(U,V\right)$ is non-connected.
Its connected components consisting of orientation-preserving and
orientation-reversing mapping will be denoted respectively by ${\rm LI}^{+}\left(U,\omega;V,\ell\right)$ and ${\rm LI}^{-}\left(U,\omega;V,\ell\right)$. Obviously, if $U=V$, these subsets are objectively defined, do not depend on any choice of orientation standards and are simply given by the group ${\rm GL}^{+}(V)$ and its coset ${\rm GL}^{-}(V)$. The subgroup ${\rm GL}^{+}(V)$ preserves any of two possible orientations in $V$ and the coset ${\rm GL}^{-}(V)$ interchanges
them mutually. Obviously, ${\rm LI}^{+}\left(U,V\right)$, ${\rm LI}^{-}\left(U,V\right)$ are homogeneous
spaces of ${\rm GL}^{+}(V)\times{\rm GL}^{+}(U)$ acting through (\ref{eq.91}). 
Also the action of the coset ${\rm GL}^{-}(V)\times{\rm GL}^{-}(U)$ preserves separately ${\rm LI}^{+}\left(U,V\right)$, ${\rm LI}^{-}\left(U,V\right)$, whereas the cosets ${\rm GL}^{+}(V)\times{\rm GL}^{-}(U)$, ${\rm GL}^{-}(V)\times{\rm GL}^{+}(U)$ interchange them.

Similarly, the isochoric groups ${\rm UL}(V)$, ${\rm SL}(V)$, just like those
for $U$, are objectively defined, i.e., do not assume any orientation
or volume normalization, but for mappings acting between different
linear spaces $U$, $V$ it is no longer the case. Let $\mu$, $\nu$
denote the Lebesgue measures on $U$, $V$ with some fixed normalizations.
For example, they may be ones induced by the metric tensors $\eta\in U^{\ast}\otimes U^{\ast}$, $g\in V^{\ast}\otimes V^{\ast}$. The symbol ${\rm UL}(U,\mu;V,\nu)$ will be denote the set of linear isomorphisms transforming the measure $\mu$ into $\nu$. Obviously, it is also a homogeneous space of ${\rm UL}(V)\times{\rm UL}(U)$ acting through (\ref{eq.91}). And again we can distinguish the subset if isomorphisms preserving the oriented volumes: \begin{equation}
{\rm SL}(U,\mu,\omega;V,\nu,\ell):={\rm LI}^{+}\left(U,\omega;V,\ell\right)\bigcap {\rm UL}(U,\mu;V,\nu).\label{eq.110}
\end{equation}
It is of course the homogeneous space of ${\rm SL}(V)\times{\rm  SL}(U)$ acting
through (\ref{eq.91}). More precisely, it is an orbit of this transformation
group acting in ${\rm LI}\left(U,V\right)$. 

\section{Degrees of Freedom, Inertia and Kinematics of Affine Bodies}

Let us remind the basis ideas concerning affine motion of a single
body. This is necessary for formulating the system dynamics. We must
begin with some formalism, just starting from repeating the notation
from our previous paper \cite{acta,all04,all05}. Obviously, our objects move in the three-dimensional physical space; however, in many problems it is
just more convenient (even from the technical point of view) to work
in an $n$-dimensional space and only on the stage of real applications
put $n=3$ (sometimes $n=2,1$, when dealing with degenerate dimension
situations). So, let $\left(M,V,\rightarrow\right)$ denote, as in
all mentioned former papers, the physical affine space; $M$ is the
underlying point set, $V$ is the linear $n$-dimensional space of translations
in $M$, and the arrow $\rightarrow$ denotes the operation which to
every pair of points $x,y\in M$ assigns the only translation vector
$\overrightarrow{xy}\in V$ "originating" at $x\in M$ and "terminating"
at $y\in M$, i.e., carrying over $x$ into $y$. All standard axioms
of affine geometry are assumed \cite{Hel_93,acta,all04,all05}; 
let us remind them: 
\begin{itemize}
\item $\overrightarrow{xy}+\overrightarrow{yz}+\overrightarrow{zx}=0$,

\item for any $y\in M$ the mapping
$M\ni x\mapsto\overrightarrow{yx}\in V$ is a bijection of $M$ onto $V$.
\end{itemize}
These axioms imply in a standard way that 
\begin{equation}
\overrightarrow{xy}+\overrightarrow{yx}=0,\qquad \overrightarrow{xx}=0
\label{eq.113}\end{equation}
for any $x,y\in M$. The linear space $V$ as an additive Abelian
group is canonically isomorphic with the translation group $T(M)$.
This group consists of mappings 
\begin{equation}
t_{v}:M\rightarrow M,\qquad v\in V
\end{equation}
such that for any $x\in M$, $y=t_{v}(x)\in M$ is the only point for which 
\begin{equation}
\overrightarrow{xy}=v.\label{eq.114}
\end{equation}
This action of $V$ on $M$ as a transformation group is transitive,
free and effective. 

As usual, we say that $A:M\rightarrow M$ is an affine mapping
if there exists a linear mapping ${\rm L}(A):V\rightarrow V$, denoted
also for obvious reasons as $DA$, such that 
\begin{equation}
\overrightarrow{A(x)A(y)}=L(A)\overrightarrow{xy}.\label{eq.115}
\end{equation}
Such mappings preserve all affine relations, i.e., they map straight
lines into straight lines and preserve parallelism of straight lines.
It is obvious that 
\begin{equation}
L(AB)=L(A)L(B),\qquad L\left(t_{v}\right)={\rm Id}_{M}.\label{eq.116}
\end{equation}
In this way ${\rm L}$ is a homomorphism of the semigroup ${\rm Af}(M)$ of all
affine mappings of $M$ into $M$ into the semigroup ${\rm L}(V)$ of all
linear mappings of $V$ into $V$. Invertible elements of ${\rm Af}(M)$
form the group ${\rm GAf}(M)$, i.e., the group of affine transformations of $M$
onto $M$. When restricted to ${\rm GAf}(M)$, ${\rm L}$ is a group homomorphism
of ${\rm GAf}(M)$ onto ${\rm GL}(V)$. Translations are affine mappings; the
only ones for which the second of (\ref{eq.116}) holds. $T(M)$ is a normal
subgroup of ${\rm GAf}(M)$.

Fixing some reference point-origin, $\mathcal{O}\in M$, we can identify
${\rm GAf}(M)$ with the semidirect product ${\rm GL}(V)\times_{s}V$;
any mapping $\phi\in{\rm GAf}(M)$ is identified with the pair 
\begin{equation}
\left(L[\phi],\overrightarrow{\mathcal{O}\phi\left(\mathcal{O}\right)}\right).
\end{equation}
The composition of $\phi$-mappings is represented by the semidirect
rule multiplication of indicated pairs. If some fixed affine coordinates
are used, then the point $x$ with coordinates $x^{i}$ is transformed
into the point $y$ with coordinates $y^{i}$, where
\begin{equation}
y^{i}=A^{i}{}_{j}x^{j}+v^{i}\label{eq.117}
\end{equation}
("linear-nonhomogeneous function" in the school language, simply the first-degree polynomial in affine coordinates). If $\phi\in{\rm GAf}(M)$, then obviously $\det[A^{i}{}_{j}]\neq 0$; this need not be satisfied for a general $\phi\in{\rm Af}(M)$. By analogy to linear transformations one uses the concepts of unimodular and special affine subgroups ${\rm UAf}(M)$, ${\rm SAf}(M)$. They consist of such $\phi\in{\rm GAf}(M)$ that respectively 
\begin{equation}
L[\phi]\in{\rm UL}(V),\qquad L[\phi]\in{\rm SL}(V). 
\end{equation}
Similarly $\phi\in{\rm GAf}^{+}(M)$, $\phi\in{\rm GAf}^{-}(M)$, when, respectively, 
\begin{equation}
L[\phi]\in{\rm GL}^{+}(V),\qquad L[\phi]\in{\rm GL}^{-}(V). 
\end{equation}
These subgroups or cosets are defined objectively in a bare $(M,V,\rightarrow)$ without any additional structure like the volume or orientation standards in $M$ (equivalently in $V$). If $(M,V,\rightarrow)$ is structurally enriched by the metric tensor $g\in{\rm Sym}\left(V^{\ast}\otimes V^{\ast}\right)$, i.e., if we deal with the Euclidean space $(M,V,\rightarrow,g)$, then the isometry subgroups ${\rm OAf}(M,g)$, 
${\rm SOAf}(M,g)$ may be introduced as well as the coset ${\rm OAf}^{-}(M,g)$. Their elements $\phi$ satisfy, respectively, the following conditions: 
\begin{equation}
L[\phi]\in{\rm O}\left(V,g\right),\qquad L[\phi]\in{\rm SO}\left(V,g\right),\qquad L[\phi]\in{\rm O}^{-}\left(V,g\right).
\end{equation} 

The material space $N$ is also endowed with an affine structure $(N,U,\rightarrow)$ with the same meaning of symbols as above. In "realistic"
problems some metric tensor $\eta\in U^{\ast}\otimes U^{\ast}$ is fixed and one
deals with the Euclidean space, i.e., the quadruple $(N,U,\rightarrow,\eta)$.
Usually we were dealing with non-degenerate situations when $M$, $N$
had the same dimension $n$ ($U$, $V$ as linear spaces had the same dimension).
Nevertheless, there are also models where $\dim N<\dim M$ \cite{Roz_05,Roz_10}. 

Affine mappings acting between different spaces are defined just like
ones acting within the same space, i.e., preserving all affine relations.
So, $\varphi:N\rightarrow M$ is affine if there exist a linear
mapping $L[\varphi]\in{\rm L}\left(U,V\right)$ such that 
\begin{equation}
\overrightarrow{\varphi(x)\varphi(y)}=L[\varphi]\overrightarrow{xy}\label{eq.118}
\end{equation}
for any $x,y\in N$. Obviously, for the triple of affine spaces the
following chain rule holds:
\begin{equation}
L\left[\varphi_{1}\circ\varphi_{2}\right]=L[\varphi_{1}]L[\varphi_{2}].\label{eq.119}
\end{equation}
The set of all affine mappings from $N$ to $M$ will be denoted by ${\rm Af}\left(N,M\right)$ or more detailly ${\rm Af}(N,U,\rightarrow;M,V,\rightarrow)$. It is itself an affine space in a canonical way; its translation space is ${\rm Af}\left(N,M\right)$. In the last expression the linear space $V$ is interpreted as an affine space in the canonical way, namely: 
\begin{equation}
\overrightarrow{uv}:=v-u. 
\end{equation}
The Lagrange and Euler affine coordinates will be denoted by $a^{K}$, $y^{i}$
respectively. They are fixed when some origins $\mathcal{O}\in N$,
$\mathcal{O}\in M$ and linear frames $\left(\ldots,E_{K},\ldots\right)$, 
$(\ldots,e_{i},\ldots)$ are distinguished, 
\begin{equation}
\overrightarrow{\mathcal{O}y}=y^{i}e_{i},\qquad
\overrightarrow{\mathcal{O}A}=a^{K}E_{K}.\label{eq.120}
\end{equation}
Affine transformation $\varphi\in{\rm Af}\left(N,M\right)$ maps $A\in N$ with coordinates
$a^{K}$ onto $y\in M$ with coordinates $y^{i}$ such that 
\begin{equation}
y^{i}=x^{i}+\varphi^{i}{}_{K}a^{K}.\label{eq.121}
\end{equation}

These are natural coordinates $\left(x^{i},\varphi^{i}{}_{K}\right)$ on ${\rm Af}\left(N,M\right)$ induced by the choice of affine frames $(\mathcal{O};\ldots,e_{i},\ldots)$ and $(\mathcal{O};\ldots,E_{K},\ldots)$ respectively in $M$ and $N$. The submanifold of affine monomorphisms of $N$ into $M$ will be denoted by ${\rm AfM}\left(N,M\right)$. If, as it is usually the case in our models, $\dim N=\dim M=n$, then ${\rm AfM}\left(N,M\right)$ becomes the manifold of affine isomorphisms ${\rm AfI}\left(N,M\right)$. The ${\rm L}$-operations map them respectively onto manifolds ${\rm LM}\left(U,V\right)$, ${\rm LI}\left(U,V\right)$. And in analogy to submanifolds 
${\rm LI}^{+}\left(U,\omega;V,\ell\right)$, ${\rm LI}^{-}\left(U,\omega;V,\ell\right)$, subgroups ${\rm GL}^{+}(U)$, ${\rm GL}^{+}(V)$, cosets ${\rm GL}^{-}(U)$, ${\rm GL}^{-}(V)$, one uses their ${\rm L}$-co-images, i.e., submanifolds ${\rm AfI}^{+}\left(N,\omega;M,\ell\right)$, ${\rm AfI}^{-}\left(N,\omega;M,\ell\right)$, subgroups ${\rm GAf}^{+}(N)$, 
${\rm GAf}^{+}(M)$, and cosets ${\rm GAf}^{-}(N)$, ${\rm GAf}^{-}(M)$. Similarly, the symbols ${\rm SAf}\left(N,\mu,\omega;M,\nu,\ell\right)$ and ${\rm UAf}\left(N,\mu;M,\nu\right)$ are ${\rm L}$-co-images of ${\rm SL}\left(U,\mu,\omega;V,\nu,\ell\right)$ and ${\rm SL}\left(U,\mu;V,\nu\right)$. And the groups ${\rm SAf}(N)$, ${\rm SAf}(M)$, ${\rm UAf}(N)$, ${\rm UAf}(M)$ are ${\rm L}$-co-images of linear subgroups ${\rm SL}(U)$, ${\rm SL}(V)$, ${\rm UL}(U)$, ${\rm UL}(V)$. In other words, one speaks about affine mappings preserving orientations, volume or both of them simultaneously.

If metric tensors $\eta\in U^{\ast}\otimes U^{\ast}$, $g\in V^{\ast}\otimes V^{\ast}$ are taken into account, i.e., if we distinguish orthogonal subgroups ${\rm O}\left(U,\eta\right)$, ${\rm O}\left(V,g\right)$, ${\rm SO}\left(U,\eta\right)$, ${\rm SO}\left(V,g\right)$, the orientation-reversing cosets ${\rm O}^{-}\left(U,\eta\right)$, ${\rm O}^{-}\left(V,g\right)$, then in the affine groups ${\rm GAf}(N)$, ${\rm GAf}(M)$ automatically the corresponding groups or cosets of Euclidean motions are distinguished just again as the ${\rm L}$-co-images of the mentioned linear subgroups and cosets. Roughly speaking, they are isomorphic with the semidirect products of linear groups and translation groups in $N$, $M$. We shall use the symbols ${\rm O}\left(U,\eta;V,g\right)$, ${\rm E}\left(N,\eta;M,g\right)$ respectively for the manifold of linear isometries of $\left(U,\eta\right)$ onto $\left(V,g\right)$ (i.e., such mappings $\varphi\in{\rm L}\left(U,V\right)$ that $\eta=\varphi^{\ast}g$) and for the manifold of affine isometries of $\left(N,U,\rightarrow,\eta\right)$ onto $\left(M,V,\rightarrow,g\right)$; obviously, ${\rm E}\left(N,\eta;M,g\right)$ is the ${\rm L}$-co-image of ${\rm O}\left(U,\eta;V,g\right)$. Just like the total 
${\rm AfI}\left(N,M\right)$, its submanifold ${\rm E}\left(N,\eta;M,g\right)$ is non-connected and consists of two connected components. When orientations are fixed in $N$, $M$ (in $U$, $V$), those components consist respectively of the orientation-preserving and orientation-reversing mappings (the corresponding $\varphi=L\left[\varphi\right]$ are orientation-preserving or orientation-reversing linear mappings).

The configuration space $Q$ of the affinely-rigid body is simply given
by the manifold ${\rm AfI}\left(N,M\right)$ or rather by one of its two connected
components. When orientations $\omega$, $\ell$ are fixed in $N$, $M$ (in
$U$, $V$), $Q$ is identified with ${\rm AfI}^{+}\left(N,\omega;M,\ell\right)$, the submanifold of orientation-preserving mappings. Strictly speaking, the restriction to connected submanifolds is essential for classical continuous bodies, not necessarily for discrete ones. And the more so in quantized models the non-connected configuration spaces seem to be admissible. If some
reference point $\mathcal{O}\in N$ is fixed in the material space,
then $Q$ may be naturally identified with the Cartesian product 
\begin{equation}
Q\simeq Q_{\rm tr}\times Q_{\rm int}=M\times {\rm LI}\left(U,V\right);\label{eq.122}
\end{equation}
the labels "tr", "int" refer respectively to the translational and internal/relative motions. Just as previously, in traditional problems one restricts to 
\begin{equation}
Q\simeq M\times{\rm LI}^{+}\left(U,\omega;V,\ell\right).\label{eq.123}
\end{equation}

Usually (not always) the reference point $\mathcal{O}$ is chosen
as the Lagrangian centre of mass. If the mass distribution in $N$
is given by the positive, time-independent measure $\mu$ on $N$,
then, by definition of the centre,
\begin{equation}
\int\overrightarrow{\mathcal{O}X}d\mu(X)=0.\label{eq.124}
\end{equation}
And the affine Lagrange coordinates $a^{K}$ in $N$ are chosen in such
a way that $a^{K}\left(\mathcal{O}\right)=0$, where $K=1,\ldots,n$.

Obviously, the identification (\ref{eq.122}) is meant in the sense that
the affine mapping $\phi:N\rightarrow M$ corresponds to the pair
$(m,\varphi)$ where 
\begin{equation}
m=\phi\left(\mathcal{O}\right)=
\mathcal{O}_{\phi},\qquad \varphi=L\left[\phi\right].\label{eq.125}
\end{equation}

If affine coordinates $a^{K}$, $y^{i}$ are fixed in $N$, $M$ then
Lagrange and Euler coordinates are interrelated by (\ref{eq.121}); $x^{i}$
are coordinates of the centre of mass $\mathcal{O}_{\phi}$ in $M$,
$\varphi^{i}{}_{K}$ are internal generalized coordinates. This
is the most natural choice of generalized coordinates in $Q$. It
happens very often that the Lagrange coordinates $a^{K}$ are fixed
once for all, often due to some physical conditions. Then the material
space $N$ is simply identified with $\mathbb{R}^{n}$. The manifold
of linear isomorphisms ${\rm LI}\left(U,V\right)$ becomes identified with the manifold of linear frames in $V$, denoted by $F(V)$. Namely, $\varphi\in{\rm LI}\left(U,V\right)$ is identified with $e=\left(\ldots,e_{K},\ldots\right)$, where $e_{K}=\varphi\varepsilon_{K}$ and $\left(\ldots,\varepsilon_{K},\ldots\right)$ is the natural basis of $\mathbb{R}^{n}$. If there is no danger of misunderstanding, $e_{K}$ are simply denoted by $\varphi_{K}$.

If affine motion is subject to additional constraints, the configuration
space becomes some submanifold of $Q$. Usually we mean constraints
imposed on internal motion. For example, if there is no deformation
and the body is rigid in the usual metrical sense, then $Q_{\rm int}$
becomes reduced to the manifold of linear isometries ${\rm O}\left(U,\eta;V,g\right)$, or rather to its connected component ${\rm O}^{+}\left(U,\omega;V,\ell\right)$.

If the body is incompressible and orientation-preserving, $Q_{\rm int}$
is reduced to ${\rm SLI}\left(U,\mu,\omega;M,\nu,\ell\right)$, i.e., 
$Q$ is reduced to ${\rm SAfI}\left(U,\mu,\omega;M,\nu,\ell\right)$.

In usual models there are two inertial objects, namely the total mass
$m$ and the second-order (quadrupole) moment of Lagrangian mass distribution:
\begin{equation}
m=\int d\mu(a),\qquad J^{AB}=\int a^{A}a^{B}d\mu(a)=J^{BA}.\label{eq.126}
\end{equation}

The scalar $m$ and the Lagrangian tensor $J\in U\otimes U$ are constant
and characterize respectively the inertia of translational and internal
(relative) motions. Obviously, one can calculate all possible moments
of this kind, 
\begin{equation}
J(k)^{A_{1},\ldots,A_{k}}=\int a^{A_{1}}\cdots a^{A_{K}}d\mu(a),\label{eq.127}
\end{equation}
however, $J(1)=0$ according to our convention concerning coordinates
$a^{K}$ (their origin in $U$ is placed at the Lagrangian centre of
mass distribution $\mu$) and high-order moments $J(k)$, $k>2$,
are irrelevant for the affine motion. The second-order moment is algebraically
equivalent to the one that is usually called the co-moving inertial tensor.

Being explicitly dependent on the configuration $\phi$, thus, on time,
the Euler moments, i.e., multipoles of the transported measure $\mu_{\phi}$
with respect to the spatial coordinates $y^{i}$ are non-useful as inertial
characteristics. This fact is known from the (metrically) rigid body
mechanics.

As usual, the motion is described by the time-dependence of the configuration
$\phi$, i.e., by curves 
\begin{equation}
\mathbb{R}\ni t\mapsto\phi(t)\in Q,
\end{equation}
or equivalently, in terms of generalized coordinates 
\begin{equation}
\mathbb{R}\ni t\mapsto x^{i}(t),\qquad
\mathbb{R}\ni t\mapsto\varphi^{i}{}_{K}(t),\qquad 
\mathbb{R}\ni t\mapsto e^{i}{}_{K}(t).
\end{equation}
Generalized velocities are denoted by $v^{i}$, $V^{i}{}_{K}$:
\begin{equation}
v^{i}(t)=\frac{dx^{i}}{dt},\qquad V^{i}{}_{K}(t)=\frac{d\varphi^{i}{}_{K}(t)}{dt}.\label{eq.128}
\end{equation}
Obviously, $v\in V$, $V\in{\rm L}\left(U,V\right)$. Often one uses also
Lie-algebraic quantities $\Omega\in{\rm L}(V)\simeq{\rm GL}(V)^{\prime}$, $\widehat{\Omega}\in{\rm L}(U)\simeq{\rm GL}(U)^{\prime}$:
\begin{equation}
\Omega=\frac{d\varphi}{dt}\varphi^{-1},\qquad
\widehat{\Omega}=\varphi^{-1}\frac{d\varphi}{dt}=
\varphi^{-1}\Omega\varphi,\label{eq.129}
\end{equation}
i.e., analytically 
\begin{equation}
\Omega^{i}{}_{j}=\frac{d\varphi^{i}{}_{A}}{dt}\varphi^{-1A}{}_{j},\qquad
\widehat{\Omega}^{A}{}_{B}=\varphi^{-1A}{}_{i}\frac{d\varphi^{i}{}_{B}}{dt}.
\label{eq.130}
\end{equation}
Roughly speaking, $\widehat{\Omega}$ is the co-moving representation
of $\Omega$. If the motion is metrically rigid, the above objects
become respectively $g$- and $\eta$-skew-symmetric, i.e., elements of Lie algebras ${\rm SO}\left(V,g\right)^{\prime}$, 
${\rm SO}\left(U,\eta\right)^{\prime}$ of orthogonal groups 
${\rm SO}\left(V,g\right)$, ${\rm SO}\left(U,\eta\right)$:
\begin{equation}
\Omega^{i}{}_{j}=-\Omega_{j}{}^{i}=-g_{ja}g^{ib}\Omega^{a}{}_{b},\qquad
\widehat{\Omega}^{A}{}_{B}=-\widehat{\Omega}_{B}{}^{A}=
-\eta_{BK}\eta^{AL}\widehat{\Omega}^{K}{}_{L}.\label{eq.131}
\end{equation}
This is then the angular velocity respectively in laboratory and co-moving
representations. For the general affine motion one can introduce the
$g$- and $\eta$-skew-symmetric parts of $\Omega$, $\widehat{\Omega}$:
\begin{equation}
\omega^{i}{}_{j}:=\frac{1}{2}\left(\Omega^{i}{}_{j}-\Omega_{j}{}^{i}\right),\qquad
\widetilde{\omega}^{A}{}_{B}:=\frac{1}{2}\left(\widehat{\Omega}^{A}{}_{B}-
\widehat{\Omega}_{B}{}^{A}\right).\label{eq.132}
\end{equation}
This notation is not very happy, because if $\varphi$ is not an isometry,
then $\widetilde{\omega}^{A}{}_{B}$ are not co-moving components of $\omega$:
\begin{equation}
\widetilde{\omega}^{A}{}_{B}\neq\varphi^{-1A}{}_{j}\omega^{i}{}_{j}
\varphi^{j}{}_{B}.\label{eq.133}
\end{equation}

Obviously, if motion is metrically rigid, 
$\varphi\in{\rm LI}\left(U,\eta;V,g\right)$, then 
\begin{equation}
\omega^{i}{}_{j}=\Omega^{i}{}_{j},\qquad
\widetilde{\omega}^{A}{}_{B}=\widehat{\Omega}^{A}{}_{B}=
\varphi^{-1A}{}_{i}\omega^{i}{}_{j}\varphi^{j}{}_{B}.\label{eq.134}
\end{equation}
Similarly, one introduces distortional velocities as symmetric parts
of $\Omega$-objects:
\begin{equation}
d^{i}{}_{j}:=\frac{1}{2}\left(\Omega^{i}{}_{j}+\Omega_{j}{}^{i}\right),\qquad
\widetilde{d}^{A}{}_{B}:=\frac{1}{2}\left(\widehat{\Omega}^{A}{}_{B}+
\widehat{\Omega}_{B}{}^{A}\right).\label{eq.135}
\end{equation}
In rigid motion they vanish, of course, and in deformative motion,
$\widetilde{d}^{A}{}_{B}$ again are not co-moving components of $d^{i}{}_{j}$.
Let us introduce $\widehat{d}\in U^{\ast}\otimes U^{\ast}$ given by \begin{equation}
\widehat{d}_{AB}=\eta_{AC}\widetilde{d}^{C}{}_{B}.\label{eq.136}
\end{equation}
One can show that 
\begin{equation}
\widehat{d}_{AB}=\frac{1}{2}\frac{dG_{AB}}{dt}.\label{eq.137}
\end{equation}

The quantity $\Omega^{i}{}_{j}$ occurs in Eringen's papers under the
name "gyration". It is directly related to the gradient of the Euler velocity field. Indeed, if our affinely-deformable continuum fills the whole physical space $M$, then the velocity field in $M$ is given by
\begin{equation}
{\rm v}(y)^{i}=v^{i}+\Omega^{i}{}_{j}\left(y^{j}-x^{j}\right)=
\frac{dx^{i}}{dt}+\Omega^{i}{}_{j}\left(y^{j}-x^{j}\right).\label{eq.138}
\end{equation}

In certain formulas it is convenient to use the co-moving representation
of translational velocity, $\widehat{v}\in U$: 
\begin{equation}
\widehat{v}=\varphi^{-1}v,\qquad
\widehat{v}^{A}=\varphi^{-1A}{}_{i}v^{i}.\label{eq.139}
\end{equation}
Transformations (\ref{eq.91}) affect the affine velocities $\Omega$, $\widehat{\Omega}$ according to the rule: 
\begin{equation}
\varphi\mapsto A\varphi B:\qquad \Omega\mapsto A\Omega A^{-1},\qquad \widehat{\Omega}\mapsto B^{-1}\widehat{\Omega}B
\label{eq.140}
\end{equation}
for any $A\in{\rm GL}(V)$, $B\in{\rm GL}(U)$, and $\varphi\in{\rm LI}\left(U,V\right)$.

Similarly, the translational velocity transforms then as follows:
\begin{equation}
\varphi\mapsto A\varphi B:\qquad v\mapsto Av,\qquad
\widehat{v}\mapsto B^{-1}\widehat{v}.\label{eq.141}
\end{equation}

Let us mention some important point. The holonomic rigid body constraints
may be alternatively described in an apparently non-holonomic form,
just by stating that affine velocity is skew-symmetric in the $g$-sense:
\begin{equation}
\Omega^{i}{}_{j}+\Omega_{j}{}^{i}=
\Omega^{i}{}_{j}+g^{ia}g_{jb}\Omega^{b}{}_{a}=0\label{eq.142}
\end{equation}
Expression on the left-hand side is a combination of generalized
velocities $V^{i}{}_{A}=d\varphi^{i}{}_{A}/dt$ with $\varphi$-dependent
coefficients. Nevertheless, the corresponding Pfaff problem is integrable
and the family of integral surfaces is labelled by metrics $\eta\in U^{\ast}\otimes U^{\ast}$ and consists of submanifolds ${\rm O}\left(U,\eta;V,g\right)\subset{\rm LI}\left(U,V\right)$. Obviously,
in realistic problems $\eta$ is fixed once for all and only one of
all possible integral "leaves" is used as gyroscopic constraints. 

Alternatively, one can use the co-moving nonholonomic description in which $\widehat{\Omega}$ is $\eta$-skew-symmetric: 
\begin{equation}
\widehat{\Omega}^{A}{}_{B}+\widehat{\Omega}_{B}{}^{A}=
\widehat{\Omega}^{A}{}_{B}+\eta^{AC}\eta_{BD}\widehat{\Omega}^{D}{}_{C}=0
\label{eq.143}
\end{equation}
Again the Pfaff problem is completely integrable and the corresponding
integral foliation is labelled by metrics $g\in V^{\ast}\otimes V^{\ast}$ and
consists again of submanifolds ${\rm O}\left(U,\eta;V,g\right)\subset{\rm LI}\left(U,V\right)$. In "true" physics $g$ is fixed once for all, pre-established, and one obtains the same results as that based on (\ref{eq.142}).

In certain traditional books, constraints described according to the
pattern like (\ref{eq.142}), (\ref{eq.143}) are called "semi-holonomic". Strictly speaking, they give some integral foliation and no transition between different leaves is possible. Some particular leaf is fixed by physical conditions and this leaf just represents true holonomic constraints.

In a quite similar way the incompressibility constraints may be described
in the following semi-holonomic form: 
\begin{equation}
{\rm Tr}\,\Omega={\rm Tr}\,\widehat{\Omega}=0\label{eq.144}
\end{equation}

The integral foliation is labelled by the possible volume standards
$\mu$, $\nu$ in $N$, $M$ and consists of submanifolds ${\rm UL}\left(U,\mu;V,\nu\right)$ of volume-preserving isomorphisms. When $\mu$, $\nu$ are fixed (there is a canonical choice in Euclidean spaces $\left(U,\eta\right)$, $\left(V,g\right)$), then some particular leaf is fixed as holonomic constraints of incompressible motions.

In the above examples constraints (\ref{eq.142}), (\ref{eq.143}) were semi-holonomic because the subspaces of matrices (\ref{eq.142}), (\ref{eq.143}), (\ref{eq.144}) were commutator Lie algebras. 

Let us remind that the subspaces of $g$- and $\eta$-symmetric
linear mappings, i.e., ones satisfying 
\begin{eqnarray}
\Omega^{i}{}_{j}-\Omega_{j}{}^{i}=
\Omega^{i}{}_{j}-g^{ia}g_{jb}\Omega^{b}{}_{a}&=&0,\label{eq.145}\\
\widehat{\Omega}^{A}{}_{B}-\widehat{\Omega}_{B}{}^{A}=
\widehat{\Omega}^{A}{}_{B}-\eta^{AC}\eta_{BD}\widehat{\Omega}^{D}{}_{C}&=&0,
\label{eq.146}
\end{eqnarray}
are not Lie subalgebras of ${\rm L}(V)$, ${\rm L}(U)$. Moreover, they are so-to
speak as far non-subalgebras as possible; the commutators of $(g,\eta)$-symmetric mappings are just $(g,\eta)$-skew-symmetric. Because of this, constraints (\ref{eq.145}), (\ref{eq.146}) are non-holonomic, again let us say, as non-holonomic as possible. Physically we would say that they are constraints of rotation-less motion. This is the only reasonable definition of the rotation-free behaviour. As a physical application we can realize, e.g., homogeneously-deformable suspension in a viscous fluid. The friction at the surface prevents the body to rotate, but it is not an obstacle against deformations.

Let us now remind briefly the basic Hamiltonian concepts. The tangent
bundle $TQ$ of the configuration space $Q$ may be identified simply
with the Cartesian product 
\begin{equation}
\mathcal{P}_{V}=M\times{\rm LI}\left(U,V\right)\times V\times{\rm L}\left(U,V\right).\label{eq.147}
\end{equation}
Its elements $\left(x,\varphi;v,V\right)$ are quadruples consisting of the
centre of mass position $x\in M$ internal configuration $\varphi\in{\rm LI}\left(U,V\right)$, the translational velocity $v\in V$, and the generalized velocity of internal motion $V\in{\rm L}\left(U,V\right)$. But as well we can use a representation based on non-holonomic velocities, first of all the following ones: 
\begin{eqnarray}
\mathcal{P}_{\Omega}&:=&M\times{\rm LI}\left(U,V\right)\times V\times{\rm L}(V),\label{eq.148}\\
\mathcal{P}_{\widehat{\Omega}}&:=&M\times{\rm LI}\left(U,V\right)\times U\times{\rm L}(U),\label{eq.149}
\end{eqnarray}
where the elements are respectively $\left(x,\varphi;v,\Omega\right)$ and $\left(x,\varphi;\widehat{v},\widehat{\Omega}\right)$ involving, as we see, the laboratory (spatial) $\left(v,\Omega\right)$-description of velocities and their co-moving (material) $\left(\widehat{v},\widehat{\Omega}\right)$-representation.

The Hamiltonian phase space, i.e., cotangent bundle $T^{\ast}Q$, is identified
with the Cartesian product 
\begin{equation}
P_{V}:=M\times{\rm LI}\left(U,V\right)\times V^{\ast}\times{\rm L}(V,U).\label{eq.150}
\end{equation}
Its elements $\left(x,\varphi;p,P\right)$ comprise the configuration
$\left(x,\varphi\right)$ and canonical momenta $p\in V^{\ast}$ (translational
one) and $P\in{\rm L}(V,U)\simeq{\rm L}\left(U,V\right)^{\ast}$ (the internal one); one makes use here of the canonical isomorphism between ${\rm L}\left(U,V\right)^{\ast}$ and ${\rm L}\left(V,U\right)$ given by trace formula, \begin{equation}
\left\langle P,V\right\rangle={\rm Tr}\left(PV\right)=P^{A}{}_{i}V^{i}{}_{A}.\label{eq.151}
\end{equation}
And just like previously, one introduces the objects dual to affine
velocities, namely affine spin in the laboratory and co-moving representations,
$\Sigma\in{\rm L}(V)$, $\widehat{\Sigma}\in{\rm L}(U)$. They are given by \begin{equation}
\Sigma=\varphi P,\qquad
\widehat{\Sigma}=P\varphi=\varphi^{-1}\Sigma\varphi,\label{eq.152}
\end{equation}
i.e., analytically
\begin{equation}
\Sigma^{i}{}_{j}=\varphi^{i}{}_{j}P^{A}{}_{j},\qquad 
\widehat{\Sigma}^{A}{}_{B}=P^{A}{}_{i}\varphi^{i}{}_{B}.
\label{eq.153}
\end{equation}
Let us notice that the inverse mapping $\varphi^{-1}$ does not occur here, so they are defined globally on the total ${\rm L}\left(U,V\right)$. They are also well defined when $\dim U<\dim V$, but, of course, the similarity relationship in the last expression of (\ref{eq.152}) does not exist then, because it is based on the existence of $\varphi^{-1}$. So, in such a situation, $\widehat{\Sigma}$ is not a co-moving representation of $\Sigma$.

The manifolds of quadruples 
\begin{equation}
\left(x,\varphi;p,P\right),\qquad \left(x,\varphi;p,\Sigma\right),\qquad \left(x,\varphi;\widehat{p},\widehat{\Sigma}\right), 
\end{equation}
i.e., various models of the Hamiltonian phase space $T^{\ast}Q$ will be denoted respectively by $P$, $P_{\Sigma}$, $P_{\widehat{\Sigma}}$.

Affine spin is in a sense conjugate momentum of non-holonomic velocities, because
\begin{equation}
{\rm Tr}(PV)={\rm Tr}\left(\Sigma\Omega\right)={\rm Tr}\left(\widehat{\Sigma}\widehat{\Omega}\right). 
\label{eq.154}
\end{equation}
It is clear that this trace expression corresponds to the canonical isomorphism between ${\rm L}\left(U,V\right)^{\ast}$ and ${\rm L}(V,U)$, ${\rm L}(U)^{\ast}$ and ${\rm L}(U)$, ${\rm L}(V)^{\ast}$ and ${\rm L}(V)$.

Let us mention also the co-moving representation of canonical translational momentum,
\begin{equation}
U\ni\widehat{p}:=\varphi^{\ast}p=p\circ\varphi,\qquad 
\widehat{p}_{A}=p_{i}\varphi^{i}{}_{A}. \label{eq.155}
\end{equation}
As usual, the relationship between $\left(v,V\right)$ and $\left(p,P\right)$, 
or ones between $\left(v,\Omega\right)$ and $\left(p,\Sigma\right)$, $\left(\widehat{v},\widehat{\Omega}\right)$ and $\left(\widehat{p},\widehat{\Sigma}\right)$, is a dynamical concept depending on the Lagrangian (in usual dynamical models $L=T-V(x,\varphi)$ --- just on $T$). Obviously, $\left(p,\Sigma\right)$ and $\left(\widehat{p},\widehat{\Sigma}\right)$ are respectively Hamiltonian generators, i.e., momentum mappings \cite{Abr-Mar} 
of ${\rm GAf}(M)$ and ${\rm GAf}(N)$ acting on $Q={\rm LI}\left(U,V\right)$ through the left and right superpositions,
\begin{equation}
\phi\mapsto A \circ \phi \circ B, \label{eq.156}
\end{equation}
where 
\begin{equation}
(A,B)\in{\rm GAf}(M)\times GAf(N),\qquad A\in{\rm GAf}(M),\qquad B\in{\rm GAf}(N).
\end{equation}
Therefore, $p_{i}$, $\widehat{p}_{A}$ are respectively basic generators of spatial and material translations. $P^{A}{}_{i}$ are basic Hamiltonian generators of local additive translations in ${\rm LI}\left(U,V\right)$. And $\Sigma^{i}{}_{j}$, $\widehat{\Sigma}^{A}{}_{B}$ are basic Hamiltonian generators of the left and right multiplicative translations (\ref{eq.91}) in ${\rm LI}\left(U,V\right)$, i.e., spatial and material rotations and homogeneous deformations. Let us remind that just like in the case of angular momentum one speaks about "orbital", "internal" (spin) and total angular momentum, one can, just should, use their counterparts in mechanics of affine bodies. The canonical affine "orbital" momentum with respect to some fixed "origin" $\mathcal{O}\in M$ is given by
\begin{equation}
\Lambda \left[\mathcal{O}\right]:= \overrightarrow{\mathcal{O}\mathcal{O}_{\phi}}\otimes p= \overrightarrow{\mathcal{O}x}\otimes p\in V \otimes V^{\ast}\simeq {\rm L}(V),\label{eq.157}
\end{equation}
i.e., analytically, simply
\begin{equation}
\Lambda \left[\mathcal{O}\right]^{i}{}_{j}=x^{i}p_{j}. 
\label{eq.158}
\end{equation}

The total affine momentum with respect to $\mathcal{O}\in M$ is given by
\begin{equation}
\mathcal{J}\left[\mathcal{O}\right]:=\Lambda\left[\mathcal{O}\right]+\Sigma. \label{eq.159}
\end{equation}
It is a system of Hamiltonian generators of centro-affine mappings, i.e., of the subgroup ${\rm GAf}\left(M,\mathcal{O}\right)\subset{\rm GAf}(M)$ preserving $\mathcal{O}\in M$ and acting on the left on configurations via (\ref{eq.156}). The $g$-skew-symmetric parts of $\mathcal{J}\left[\mathcal{O}\right]$, $\Lambda \left[\mathcal{O}\right]$, $\Sigma$ are respectively the total, orbital, and internal (spin) angular momenta. They generate the left (spatial) action of the Euclidean subgroup ${\rm E}(M,g)\subset{\rm Af}(M)$ operating via (\ref{eq.156}). In the velocity representation, via the inverse Legendre transformations, those all quantities have to do with the dipole moments of the distribution of kinematical linear momentum. The $\eta$-skew-symmetric part of $\widehat{\Sigma}$ was by Dyson called "vorticity" \cite{Dys_68}
\begin{equation}
V^{A}{}_{B}:=\widehat{\Sigma}^{A}{}_{B}-\widehat{\Sigma}_{B}{}^{A}= \widehat{\Sigma}^{A}{}_{B}-\eta^{AC}\eta_{BD}\widehat{\Sigma}^{D}{}_{C}. \label{eq.160}
\end{equation}
{\bf Remark:} $\widehat{V}^{A}{}_{B}$ is not the co-moving representation of spin
\begin{equation}
S^{i}{}_{j}:=\Sigma^{i}{}_{j}-\Sigma_{j}{}^{i}= 
\Sigma^{i}{}_{j}-g^{ik}g_{jl}\Sigma^{l}{}_{k}, 
\label{eq.161}
\end{equation}
i.e.,
\begin{equation}
S^{i}{}_{j}\neq \varphi^{i}{}_{A}V^{A}{}_{B}\varphi^{-1B}{}_{j},
\label{eq.162}
\end{equation}
unless the motion is metrically rigid, $\varphi\in{\rm O}\left(U,\eta;V,g\right)$.

Transformation rules of $\Sigma$, $\widehat{\Sigma}$ are identical with those of $\Omega$, $\widehat{\Omega}$ (\ref{eq.140}), thus,,
\begin{equation}
\varphi\mapsto A\varphi B:\qquad \Sigma\mapsto A\Sigma A^{-1},\qquad \widehat{\Sigma}\mapsto B^{-1}\widehat{\Sigma}B
\label{eq.163}
\end{equation}
for any $A\in{\rm GL}(V)$, $B\in{\rm GL}(U)$.

As usual, the most convenient way of deriving equations of motion for Lagrangian-Hamiltonian models is to use the Poisson brackets for the basic dynamical quantities and write down equations in the following form:
\begin{equation}
\frac{dF}{dt}=\left\{F,H\right\},\label{eq.164}
\end{equation}
where $F$ runs over some family of functionally independent functions on the phase space. Usually this family includes the above quantities. The latter ones satisfy the Poisson constants rules discussed in \cite{acta,all04,all05} and reflecting the affine group structure constants.

Let us remind some of them:
\begin{eqnarray}
\{\Sigma^{i}{}_{j},\Sigma^{k}{}_{l}\}&=&\delta^{i}{}_{l}\Sigma^{k}{}_{j}-
\delta^{k}{}_{j}\Sigma^{i}{}_{l},\label{eq.164b1}\\
\{\Lambda^{i}{}_{j},\Lambda^{k}{}_{l}\}&=&\delta^{i}{}_{l}\Lambda^{k}{}_{j}-
\delta^{k}{}_{j}\Lambda^{i}{}_{l},\label{eq.164b2}\\
\{\mathcal{J}^{i}{}_{j},\mathcal{J}^{k}{}_{l}\}&=&
\delta^{i}{}_{l}\mathcal{J}^{k}{}_{j}-
\delta^{k}{}_{j}\mathcal{J}^{i}{}_{l},\label{eq.164b3}\\
\left\{\widehat{\Sigma}^{A}{}_{B},\widehat{\Sigma}^{C}{}_{D}\right\}&=&
\delta^{C}{}_{B}\widehat{\Sigma}^{A}{}_{D}-
\delta^{A}{}_{D}\widehat{\Sigma}^{C}{}_{B},\label{eq.164b4}\\
\left\{\Sigma^{i}{}_{j},\widehat{\Sigma}^{C}{}_{D}\right\}&=&0,
\label{eq.164b5}\\
\left\{\widehat{\Sigma}^{A}{}_{B},\widehat{p}_{C}\right\}&=&
\delta^{A}{}_{C}\widehat{p}_{B},\label{eq.164b6}\\
\{\mathcal{J}^{i}{}_{j},p_{k}\}=\{\Lambda^{i}{}_{j},p_{k}\}&=&
\delta^{i}{}_{k}p_{j}.\label{eq.164b7}
\end{eqnarray}
If $F$ is any function depending only on generalized coordinates $x^{i}$, $\varphi^{i}{}_{A}$, then
\begin{eqnarray}
\{F,\Sigma^{i}{}_{j}\}&=&\varphi^{i}{}_{A}\frac{\partial F}{\partial \varphi^{j}{}_{A}},\label{eq.164b8}\\
\{F,\Lambda^{i}{}_{j}\}&=&x^{i}\frac{\partial F}{\partial x^{j}},\label{eq.164b9}\\
\{F,\widehat{\Sigma}^{A}{}_{B}\}&=&\varphi^{i}{}_{B}\frac{\partial F}{\partial \varphi^{i}{}_{A}}.\label{eq.164b10}
\end{eqnarray}

\section{Towards the Dynamics of Systems of Affine Bodies}

Let us now discuss some peculiarities of the system of affine bodies. Certain quite new problems appear there, especially, ones concerning the dynamical affine invariance. In mechanics of a single affine body there exist affinely-invariant models of the kinetic energy, i.e., affinely-invariant (under left or right regular translations) metric tensors ${\rm AfI}\left(N,M\right)$. But of course there are no affinely invariant models of the potential energy, because the groups ${\rm GAf}(N)$, ${\rm GAf}(M)$ act transitively on ${\rm AfI}\left(N,M\right)$ and the only invariant functions are constant ones. However, potentials of mutual interactions between different affine bodies MAY BE affinely invariant. In any case there exist affinely-invariant scalar functions on the configuration space of the system of affine bodies. This again raises the question as to the applicability of such affine scalars as arguments of the total potential energy. Scalars of this kind were constructed above; take e.g., (\ref{eq.104}), and the question arises as to what extent, if any at all, they may be used as arguments of the potential energy, perhaps as ones in addition to the well-established orthogonal invariants $\mathcal{K}_{a}$ (\ref{eq.106}).

Let us now consider the system of $\mathcal{N}$ affine bodies. To be not too abstract let us think that those bodies are molecules of molecular crystals, colliding fullerens, inclusions or gas bubbles in fluids, or some finite elements, e.g., in solids.

The configuration space $Q\left(\mathcal{N}\right)$ of such a system is the Cartesian product, $\mathcal{N}$-th Cartesian power of $Q \simeq M \times{\rm LI}\left(U,V\right)$, the configuration space of an individual affine object:
\begin{equation}
Q\left(\mathcal{N}\right)=Q^{\mathcal{N}}=\times_{\mathcal{N}}Q. 
\label{eq.165}
\end{equation}
Depending on what is more convenient, we can use alternative representations:
\begin{equation}
Q\left(\mathcal{N}\right)\simeq Q_{\rm tr}\left(\mathcal{N}\right)\times 
Q_{\rm int}\left(\mathcal{N}\right)=M^{\mathcal{N}}\times{\rm LI}\left(U,V\right)^{\mathcal{N}}\simeq M^{\mathcal{N}}\times F(V)^{\mathcal{N}},\label{eq.166}
\end{equation}
with the same provisos as previously that for classical continuous systems it is rather connected component of the above manifold that should be taken as the configuration space. However, even in the case of such objects like molecules this restriction may be too strong.

In this way, configuration of systems of affine bodies (let us keep in mind something like molecular crystals, or finite elements to be more concrete) are described by arrays 
\begin{equation}
\left(\ldots,x\left(\mathcal{K}\right),\ldots;
\ldots,\varphi\left(\mathcal{K}\right),\ldots\right), 
\end{equation}
where 
$\mathcal{K}=1,\ldots,\mathcal{N}$, $x\left(\mathcal{K}\right)\in
M$, and $\varphi\left(\mathcal{K}\right)\in{\rm LI}\left(U,V\right)$ or, equivalently, 
\begin{equation}
\left(\ldots,x\left(\mathcal{K}\right), \varphi\left(\mathcal{K}\right),\ldots\right) 
\end{equation}
depending on whether we order separately translational positions and internal/relative variables or we prefer to order one by one the total configurations of separate objects. Similarly, we shall use the Newton state space 
\begin{equation}
\mathcal{P}_{V}\left(\mathcal{N}\right),\qquad \mathcal{P}_{\Omega}\left(\mathcal{N}\right),\qquad \mathcal{P}_{\widehat{\Omega}}\left(\mathcal{N}\right) 
\end{equation}
and the Hamiltonian phase spaces
\begin{equation}
P_{V}\left(\mathcal{N}\right),\qquad
P_{\Omega}\left(\mathcal{N}\right),\qquad P_{\widehat{\Omega}}\left(\mathcal{N}\right). 
\end{equation}
They consist respectively of the ordered arrays of following objects:
\begin{eqnarray}
\left(x\left(\mathcal{K}\right),\varphi\left(\mathcal{K}\right); v\left(\mathcal{K}\right),V\left(\mathcal{K}\right)\right),&\quad& 
\left(x\left(\mathcal{K}\right),\varphi\left(\mathcal{K}\right);
v\left(\mathcal{K}\right),\Omega\left(\mathcal{K}\right)\right), \\
\left(x\left(\mathcal{K}\right),\varphi\left(\mathcal{K}\right); \widehat{v}\left(\mathcal{K}\right),\widehat{\Omega}\left(\mathcal{K}\right)\right),
&\quad& \left(x\left(\mathcal{K}\right),\varphi\left(\mathcal{K}\right); p\left(\mathcal{K}\right),P\left(\mathcal{K}\right)\right),\\
\left(x\left(\mathcal{K}\right),\varphi\left(\mathcal{K}\right); p\left(\mathcal{K}\right),\Sigma\left(\mathcal{K}\right)\right),&\quad& \left(x\left(\mathcal{K}\right),\varphi\left(\mathcal{K}\right);
\widehat{p}\left(\mathcal{K}\right),\widehat{\Sigma}\left(\mathcal{K}\right)
\right),
\end{eqnarray}
where $\mathcal{K}=1,\ldots,\mathcal{N}$ and, obviously,
\begin{equation}
v\left(\mathcal{K}\right),\ V\left(\mathcal{K}\right),\
\Omega\left(\mathcal{K}\right),\ \widehat{v}\left(\mathcal{K}\right),\ \widehat{\Omega}\left(\mathcal{K}\right),\ p\left(\mathcal{K}\right),\ P\left(\mathcal{K}\right),\ \Sigma\left(\mathcal{K}\right),\
\widehat{p}\left(\mathcal{K}\right),\ \widehat{\Sigma}\left(\mathcal{K}\right) 
\end{equation}
are just the previously introduced objects labelled now, in addition, by the index $\mathcal{K}$ referring to individual "particles".

So, for example,
\begin{eqnarray}
\mathcal{P}_{V}\left(\mathcal{N}\right)&=& M^{\mathcal{N}}\times
{\rm LI}\left(U,V\right)^{\mathcal{N}}\times V^{\mathcal{N}}\times 
{\rm L}\left(U,V\right)^{\mathcal{N}} \nonumber\\
&\simeq& M^{\mathcal{N}}\times V^{\mathcal{N}}\times
{\rm LI}\left(U,V\right)^{\mathcal{N}}\times
{\rm L}\left(U,V\right)^{\mathcal{N}}\simeq \mathcal{P}_{V}^{\mathcal{N}},\label{eq.167}\\
P_{P}\left(\mathcal{N}\right)&=& M^{\mathcal{N}}\times
{\rm LI}\left(U,V\right)^{\mathcal{N}}\times V^{\ast\mathcal{N}}\times
{\rm L}\left(V,U\right)^{\mathcal{N}} \nonumber\\
&\simeq& M^{\mathcal{N}}\times V^{\ast\mathcal{N}}\times
{\rm LI}\left(U,V\right)^{\mathcal{N}}\times
{\rm L}\left(V,U\right)^{\mathcal{N}}\simeq P_{P}^{\mathcal{N}},
\label{eq.168}\\
\mathcal{P}_{\Omega}\left(\mathcal{N}\right)&=& 
M^{\mathcal{N}}\times{\rm LI}\left(U,V\right)^{\mathcal{N}}\times
V^{\mathcal{N}}\times{\rm L}(V)^{\mathcal{N}}\nonumber\\
&\simeq& M^{\mathcal{N}}\times V^{\mathcal{N}}\times
{\rm LI}\left(U,V\right)^{\mathcal{N}}\times{\rm L}(V)^{\mathcal{N}}\simeq \mathcal{P}_{\Omega}^{\mathcal{N}},\label{eq.169}\\
P_{\Sigma}\left(\mathcal{N}\right)&=&
M^{\mathcal{N}}\times{\rm LI}\left(U,V\right)^{\mathcal{N}}\times 
V^{\ast\mathcal{N}}\times{\rm L}(V)^{\mathcal{N}}\nonumber\\
&\simeq& M^{\mathcal{N}}\times V^{\ast\mathcal{N}}\times
{\rm LI}\left(U,V\right)^{\mathcal{N}}\times{\rm L}(V)^{\mathcal{N}}\simeq P_{\Sigma}^{\mathcal{N}},\label{eq.170}\\
\mathcal{P}_{\widehat{\Omega}}\left(\mathcal{N}\right)&=& 
M^{\mathcal{N}}\times{\rm LI}\left(U,V\right)^{\mathcal{N}}\times
V^{\mathcal{N}}\times{\rm L}(U)^{\mathcal{N}}\nonumber\\
&\simeq& M^{\mathcal{N}}\times V^{\mathcal{N}}\times
{\rm LI}\left(U,V\right)^{\mathcal{N}}\times{\rm L}(U)^{\mathcal{N}}\simeq \mathcal{P}_{\widehat{\Omega}}^{\mathcal{N}},\label{eq.171}\\
P_{\widehat{\Sigma}}\left(\mathcal{N}\right)&=& 
M^{\mathcal{N}}\times{\rm LI}\left(U,V\right)^{\mathcal{N}}\times 
V^{\ast\mathcal{N}}\times{\rm L}(U)^{\mathcal{N}}\nonumber\\
&\simeq& M^{\mathcal{N}}\times V^{\ast\mathcal{N}}\times
{\rm LI}\left(U,V\right)^{\mathcal{N}}\times{\rm L}(U)^{\mathcal{N}}\simeq P_{\widehat{\Sigma}}^{\mathcal{N}}.\label{eq.172}
\end{eqnarray}

It is rather dull to write (the more to read) about the long list
of canonical identifications. Nevertheless, sometimes this care
about formal details (e.g., to make a distinction between $V$ and
$U$) is convenient and prevents us from doing mistakes. Concerning
the above particular details: there is some delicate point, not
taken here into account, but perhaps important in certain
problems. It is not quite clear, namely, if really always all
constituents should be described with the use of the same material
space $U$. Perhaps for any $\mathcal{K}$-th constituent its own
Lagrangian space $U\left(\mathcal{K}\right)$ should be used? Here we do not
go into such details and assume all $U\left(\mathcal{K}\right)$-s to be
identical with some standard $U$. In any case, the aforementioned
identification of $U$ for a single object with $\mathbb{R}^{n}$
(physically $\mathbb{R}^{3}$) justifies the use of a common $U$,
just $\mathbb{R}^{n}$, for all affine constituents of the
considered body. There are, however, some delicate points left.
Namely, analysis of symmetries is an important part of our study.
The groups ${\rm GL}(V)$, ${\rm GL}(U)$ act in an obvious way, on the left and
on the right, respectively, on the manifold of internal degrees of
freedom $Q_{\rm int}\simeq{\rm LI}\left(U,V\right)\simeq F(V)$ of a single object. By analogy, any $A\in{\rm GL}(V)$, $B\in{\rm GL}(U)$ acts on
$Q_{\rm int}\left(\mathcal{N}\right)\simeq{\rm LI}\left(U,V\right)^{\mathcal{N}}$ separately on all factors of the Cartesian product. But perhaps we should take the arrays 
\begin{equation}
\left(\ldots,A\left(\mathcal{K}\right),\ldots\right)\in
{\rm GL}(V)^{\mathcal{N}},\qquad 
\left(\ldots,B\left(\mathcal{K}\right),\ldots\right)\in
{\rm GL}(U)^{\mathcal{N}}, 
\end{equation}
and let them to act on $Q_{\rm int}\left(\mathcal{N}\right)\simeq{\rm LI}\left(U,V\right)^{\mathcal{N}}$ so that each $A\left(\mathcal{K}\right)$, $B\left(\mathcal{K}\right)$ act separately on any $\varphi\left(\mathcal{K}\right)\in{\rm LI}\left(U,V\right)$? And then the idea of taking independent, different $U\left(\mathcal{K}\right)$-spaces for the
$\mathcal{K}$-th objects would be reasonable. May such transformations have some physical meaning? It is still an open question. Let us notice, incidentally, that taking various transformations for various $\mathcal{K}$-s would be a kind of discrete gauge idea!

Let us now discuss Hamiltonian (and also non-Hamiltonian) dynamical
models of multibody affine systems and the problem of their affine
invariance. The simplest and most viable models are potential ones,
when Lagrangians have the following form:
\begin{equation}
L=T-V,\label{eq.173}
\end{equation}
where $T$ is the kinetic energy quadratic in generalized
velocities and $V$ depends only on the configuration, i.e., on
generalized coordinates. Geometrically $T$ is equivalent to some
Riemannian structure on the configuration space. The corresponding
Hamiltonians are obtained in a standard way via the Legendre
transformation and have the form of phase-space functions,
\begin{equation}
H=\mathcal{T}+V,\label{eq.174}
\end{equation}
where $\mathcal{T}$ is the kinetic Hamiltonian. If in some generalized
coordinates $q^{\mu}$ on the configuration space $T$ is given by
\begin{equation}
T=\frac{1}{2}\Gamma_{\alpha\beta}\frac{dq^{\alpha}}{dt}\frac{dq^{\beta}}{dt}, \label{eq.175}
\end{equation}
then
\begin{equation}
\mathcal{T}=\frac{1}{2}\Gamma^{\alpha\beta}p_{\alpha}p_{\beta}, 
\label{eq.176}
\end{equation}
where $\Gamma_{\alpha\beta}$, usually depending on $q^{\mu}$, are
interpreted as covariant metric coefficients, and $\Gamma^{\alpha\beta}$ are coefficients of its contravariant inverse,
\begin{equation}
\Gamma^{\alpha\mu}\Gamma_{\mu\beta}=\delta^{\alpha}{}_{\beta}.\label{eq.177}
\end{equation}

The quantities $p_{\alpha}$ are canonical momenta conjugate to
coordinates $q^{\alpha}$ (more precisely, to generalized velocities
$\dot{q}^{\alpha}$). The Legendre transformation for Lagrangians
(\ref{eq.173}) reads
\begin{equation}
p_{\alpha}=\frac{\partial T}{\partial\dot{q}^{\alpha}}= 
\Gamma_{\alpha\beta}\frac{dq^{\beta}}{dt}.\label{eq.178}
\end{equation}

We did not take into account forces depending on velocities, e.g.,
magnetic ones. For general models we have obviously the standard
rule:
\begin{equation}
p_{\alpha}=\frac{\partial L}{\partial\dot{q}^{\alpha}}.\label{eq.179}
\end{equation}

Even if one admits dissipative forces like the viscous friction,
the first step is always to start from Lagrangian model with its
main kinetic term $T$. Roughly speaking, this term characterizes
inertial properties of the object. Later on one introduces forces
of variational structure, i.e., derivable from some potential
$V(q)$ or generalized potential $V\left(q,\dot{q}\right)$ depending also on velocities $dq^{\alpha}/dt$. Generalized potentials may describe, e.g., magnetic or gyroscopic forces. And the final step, after deriving Euler-Lagrange
equations corresponding to the Lagrangian (\ref{eq.173}) is to
introduce, either phenomenologically or on the basis of
statistical-mechanical considerations, some essentially
non-Lagrangian, dissipative forces $\mathcal{D}_{\mu}$; the
corresponding equations of motion have the well-known form:
\begin{equation}
\frac{D}{Dt}\frac{\partial L}{\partial\dot{q}^{\mu}}-
\frac{\partial L}{\partial q^{\mu}}=\mathcal{D}_{\mu}.
\label{eq.180}
\end{equation}

The problem of the main inertial term of nonrelativistic mechanics,
\begin{equation}
\frac{D}{Dt}\frac{\partial T}{\partial\dot{q}^{\mu}}
\label{eq.181}
\end{equation}
was discussed in its various aspects in \cite{acta,all04,all05} in the theory of
a single affine body, certain more general aspects were also
discussed by Capriz \cite{Capr_89,Cap-Mar}.

It is not only theoretically interesting, but very often just
computationally effective to use Hamiltonian formalism. One
performs the Legendre transformation (\ref{eq.179}), one takes its
inverse (in the "usual" mechanics it is assumed to exist), i.e.,
one expresses generalized velocities $\dot{q}^{\mu}$ as functions
of canonical momenta $p_{\nu}$ and generalized coordinates
$q^{\nu}$, and then one constructs Hamiltonian $H$ as a function
on the phase, i.e., on the cotangent bundle $T^{\ast}Q$ over the
configuration space $Q$; analytically speaking, $H$ depends on
$q^{\mu}$, $p_{\mu}$, and perhaps explicitly on time $t$ (if $L$
does so). Let us remind that $H$ is given by the classical formula:
\begin{equation}
H\left(\ldots,q^{\mu},\ldots;\ldots,p_{\mu},\ldots\right)=
E\left(\ldots,q^{\mu},\ldots;\ldots,\dot{q}^{\mu},\ldots\right),\quad 
p_{\mu}=\frac{\partial L}{\partial\dot{q}^{\mu}}, 
\label{eq.182}
\end{equation}
where $E$ is the energy function on the tangent bundle $TQ$ locally given by
\begin{equation}
E=\dot{q}^{\mu} \frac{\partial L}{\partial \dot{q}^{\mu}}-L. \label{eq.183}
\end{equation}
As mentioned, $L$, $E$, $H$ may also depend explicitly on
time, although in the above formula we do not indicate this.
Although the non-dissipative Euler-Lagrange equations, i.e.,
(\ref{eq.180}) with $\mathcal{D}_{\mu}=0$, are traditionally
transformed into the canonical Hamilton form:
\begin{equation}
\frac{dq^{\mu}}{dt}=\frac{\partial H}{\partial p_{\mu}}, \qquad
\frac{dp_{\mu}}{dt}=-\frac{\partial H}{\partial q^{\mu}},
\label{eq.184}
\end{equation}
in many problems, both purely theoretical and computational ones,
it is more convenient to use not $q^{\mu}$, $p_{\mu}$, but some
their functions $F$ and to write equations of motion in the following form:
\begin{equation}
\frac{dF}{dt}= \left\{F,H\right\}, \label{eq.185}
\end{equation}
where the time differentiation on the left-hand side acts on the
time dependence of $F$ through $q^{\mu}$, $p_{\mu}$, and Poisson
brackets on the right-hand side are analytically expressed by the
usual formula:
\begin{equation}
\left\{F,G\right\}= \frac{\partial F}{\partial q^{\mu}}
\frac{\partial G}{\partial p_{\mu}} - \frac{\partial F}{\partial
p_{\mu}} \frac{\partial G}{\partial q^{\mu}}. \label{eq.186}
\end{equation}

The complete system of equations of motion is obtained when the
function $F$ in (\ref{eq.185}) runs over some system of $2f$
functionally independent functions on the phase space; $f$ is the
number of degrees of freedom, i.e., dimension of the configuration
space. Usually one chooses as $F$ generators of some important
groups of canonical transformations. An important technical tool
is the following property of Poisson brackets, additional one to
its Lie-algebraic properties:
\begin{equation}
\left\{f(F),G\right\}=f^{\prime}(F)\left\{F,G\right\}. 
\label{eq.187}
\end{equation}

Our generalized coordinates are functions $x\left(\mathcal{K}\right)^{i}$,
$\varphi\left(\mathcal{K}\right)^{i}{}_{A}$, whereas the corresponding canonical
momenta are denoted by $p\left(\mathcal{K}\right)_{i}$,
$P\left(\mathcal{K}\right)^{A}{}_{i}$. Similarly, with every of
$\mathcal{N}$ objects there are associated affine velocities and
affine momenta, 
\begin{equation}
\Omega\left(\mathcal{K}\right)^{i}{}_{j},\qquad
\widehat{\Omega}\left(\mathcal{K}\right)^{A}{}_{B},\qquad
\Sigma\left(\mathcal{K}\right)^{i}{}_{j},\qquad
\widehat{\Sigma}\left(\mathcal{K}\right)^{A}{}_{B}. 
\end{equation}
Quantities
corresponding to different "particles", i.e., labelled by
different $\mathcal{K}$-indices, have mutually vanishing Poisson
brackets. Obviously, the ones for a given fixed particle are
identical with the well-known expressions for a
single affine body \cite{acta,all04,all05}.

Assume that there is no mentioned "discrete gauging" and all
elements of the configuration arrays
\begin{equation}
\left(\ldots,x\left(\mathcal{K}\right),\ldots; 
\ldots,\varphi\left(\mathcal{K}\right),\ldots\right)
\end{equation}
are transformed in the sense of (\ref{eq.90a}), (\ref{eq.90a}), (\ref{eq.91}), (\ref{eq.156}) with the use of the same $\mathcal{K}$-independent mappings $A$, $B$. For example, translations $t_{v}:M \rightarrow M$ and internal mappings 
$A\in{\rm LI}(V)$, $B\in{\rm LI}(U)$ act as follows:
\begin{equation}
\left(\ldots,x\left(\mathcal{K}\right),\ldots; 
\ldots,\varphi\left(\mathcal{K}\right),\ldots\right)\mapsto 
\left(\ldots,t_{v}\left(x\left(\mathcal{K}\right)\right),\ldots; 
\ldots,A\varphi\left(\mathcal{K}\right)B,\ldots\right), 
\label{eq.188}
\end{equation}
where $v\in V$, $A\in{\rm GL}(V)$, $B\in{\rm GL}(U)$ in a given
transformation are fixed and do not depend on $\mathcal{K}$.
Geometric meaning of the phase-space quantities 
\begin{equation}
p_{i},\qquad \widehat{p}_{A},\qquad \Sigma^{i}{}_{j},\qquad
\widehat{\Sigma}^{A}{}_{B},\qquad \Lambda^{i}{}_{j},\qquad
\mathcal{J}^{i}{}_{j} 
\end{equation}
as Hamiltonian generators implies that
they are additive and that it is meaningful to introduce the total
objects characterizing the multiparticle situation:
\begin{eqnarray}
p_{i}=\sum_{\mathcal{K}}p_{i}\left(\mathcal{K}\right),&\quad& 
\widehat{p}_{A}=\sum_{\mathcal{K}}\widehat{p}_{A}\left(\mathcal{K}\right),
\label{eq.189a}\\
\Sigma^{i}{}_{j}=\sum_{\mathcal{K}}\Sigma^{i}{}_{j}\left(\mathcal{K}\right), 
&\quad& \widehat{\Sigma}^{A}{}_{B}=\sum_{\mathcal{K}} \widehat{\Sigma}^{A}{}_{B}\left(\mathcal{K}\right),\label{eq.189b}\\
\Lambda^{i}{}_{j}=\sum_{\mathcal{K}}\Lambda^{i}{}_{j}\left(\mathcal{K}\right), 
&\quad& \mathcal{J}^{i}{}_{j}=\sum_{\mathcal{K}} \mathcal{J}^{i}{}_{j}\left(\mathcal{K}\right),\label{eq.189c}
\end{eqnarray}
etc. The same concerns, e.g., the corresponding doubled
$g$-skew-symmetric parts of $\Sigma^{i}{}_{j}$,
$\Lambda^{i}{}_{j}$, $\mathcal{J}^{i}{}_{j}$, i.e., spin, orbital
angular momentum and total angular momentum. All those quantities,
just like (\ref{eq.189a})--(\ref{eq.189c}) generate global transformations in the multiparticle phase space and satisfy reasonable and geometrically well-motivated balance laws. Those become conservation laws if the
corresponding dynamical models are invariant under, e.g., (\ref{eq.188}).

Unlike (\ref{eq.189a})--(\ref{eq.189c}), the corresponding composition rules 
for kinematical quantities 
\begin{equation}
v^{i}\left(\mathcal{K}\right),\qquad
\widehat{v}^{A}\left(\mathcal{K}\right),\qquad \Omega^{i}{}_{j}\left(\mathcal{K}\right),\qquad
\widehat{\Omega}^{A}{}_{B}\left(\mathcal{K}\right), 
\end{equation}
and so on are non-additive, in general complicated, and, the most important thing,
non-objective. By this we mean that they depend in an essential
way on the dynamical model and follow from the geometric rules
(\ref{eq.189a})--(\ref{eq.189c}) via the inverse Legendre transformation.

What concerns the models of kinetic energy of a system, the most
natural idea is to assume the additivity as well, just like in
(\ref{eq.189a})--(\ref{eq.189c}). There is no logical necessity here to follow (\ref{eq.189a})--(\ref{eq.189c}), where this was a consequence of first principles. Nevertheless, it seems to be a priori a rather natural conjecture.

So, let 
\begin{equation}
T\left(\mathcal{K}\right)=T\left(x\left(\mathcal{K}\right),
\varphi\left(\mathcal{K}\right);\dot{x}\left(\mathcal{K}\right),
\dot{\varphi}\left(\mathcal{K}\right)\right) 
\end{equation}
denote the kinetic energy of
the $\left(\mathcal{K}\right)$-th particle, and 
\begin{equation}
\mathcal{T}\left(\mathcal{K}\right)=
\mathcal{T}\left( x\left(\mathcal{K}\right), \varphi\left(\mathcal{K}\right);
p\left(\mathcal{K}\right), \mathcal{P}\left(\mathcal{K}\right)\right) 
\end{equation}
be its canonical representation in the $\mathcal{K}$-th phase space,
i.e., the image of $T\left(\mathcal{K}\right)$ under the corresponding
Legendre transformation. According to the additive model we would have
\begin{equation}
T=\sum_{\mathcal{K}}T\left(\mathcal{K}\right),\qquad 
\mathcal{T}=\sum_{\mathcal{K}}\mathcal{T}\left(\mathcal{K}\right). \label{eq.190}
\end{equation}

The key problem is then the structure of one-particle contributions. In this treatment nothing was said as yet about this important problem; we merely referred to our models presented in earlier papers \cite{acta,all04,all05}. Below we remind them briefly. Before doing this let us mention, however, what would be the simplest idea competitive to (\ref{eq.190}). Obviously, it would be a binary one,
\begin{equation}
T=\frac{1}{2}\sum_{\mathcal{K},\mathcal{L}}
T\left(\mathcal{K},\mathcal{L}\right),\qquad \mathcal{T}=\frac{1}{2}\sum_{\mathcal{K},\mathcal{L}}\mathcal{T}
\left(\mathcal{K},\mathcal{L}\right),\label{eq.191}
\end{equation}
completed by the symmetry assumption:
\begin{equation}
T\left(\mathcal{K},\mathcal{L}\right)=
T(\mathcal{L},\mathcal{K}),\qquad \mathcal{T}\left(\mathcal{K},\mathcal{L}\right)=
\mathcal{T}(\mathcal{L},\mathcal{K}).
\label{eq.192}
\end{equation}

The additive model (\ref{eq.190}) is the special case of (\ref{eq.191}), namely one corresponding to
\begin{equation}
T\left(\mathcal{K},\mathcal{L}\right)=T\left(\mathcal{K}\right)
\delta_{\mathcal{K}\mathcal{L}},\qquad 
\mathcal{T}\left(\mathcal{K},\mathcal{L}\right)= \mathcal{T}\left(\mathcal{K}\right)\delta_{\mathcal{K}\mathcal{L}}. \label{eq.193}
\end{equation}

The hypothetical expression (\ref{eq.191}) in general need not be diagonal with respect to the particle labels.

Incidentally, let us mention some important interpretation
problems concerning the choice between postulates (\ref{eq.190}),
(\ref{eq.191}). In a sense this may be perhaps a semantical problem.
Namely, in fact it seems rather natural to understand the
"literal" kinetic energy in the additive form (\ref{eq.190}). But it
may be also reasonable to admit expressions like (\ref{eq.191}) as
terms of Lagrangians, interpreting, however, the non-diagonal terms
with $\mathcal{K}\neq\mathcal{L}$ ("crossing terms",
"interference terms") as contributions describing some kind of
interaction between different bodies, based on the coupling of
velocities. Let us mention that such terms may appear in certain
problems of the dynamics of systems of rigid bodies (gyroscopic
systems). In a sense, the couplings between spins resemble this
structure of interactions.

Now let us remind our earlier ideas concerning the kinetic energy of a single affine body \cite{acta,all04,all05}.

The traditional "d'Alembert" model, closest to our elementary
intuitions, is derived from the assumption that our affine body is
an aggregate of material points. Using intuitive physical terms we may realize, e.g., a "molecule" consisting of "atoms". Then, following the standard procedure of analytical mechanics, we perform the summation of elementary kinetic energies of constituents, then substitute affine constraints (homogeneous deformations), and finally obtain the classical formula
\begin{equation}
T=T_{\rm tr}+T_{\rm int}=\frac{m}{2}g_{ij}v^{i}v^{j}+
\frac{1}{2}g_{ij} V^{i}{}_{A}V^{j}{}_{B}J^{AB}. \label{eq.194}
\end{equation}
The meaning of symbols is like in (\ref{eq.121}), (\ref{eq.126}), thus, $m$ is the total mass and $J$ is the co-moving, thus, constant, tensor of inertia. The one-particle generalized coordinates $x^{i}$, $\varphi^{i}{}_{A}$ refer respectively to the spatial position of the center of mass and to the internal/relative motion. The symbols $v^{i}$, $V^{i}{}_{A}$ denote the 
corresponding generalized velocities, respectively, translational and internal ones (\ref{eq.128}). The above expression (\ref{eq.194}) is invariant under the spatial action of the group ${\rm O}\left(V,g\right)$ and the material action of ${\rm O}\left(U,J^{-1}\right)$; we mean transformations in the sense of (\ref{eq.91}). Obviously, it is also invariant under translations in $M$ and $N$. Particularly interesting is the special case of inertially isotropic body, when
\begin{equation}
J^{KL}=I\eta^{KL}, \label{eq.195}
\end{equation}
where $I$ is some positive constant. Then ${\rm O}\left(U,J^{-1}\right)=O\left(U,\eta\right)$, i.e., kinetic energy is $\eta$-isotropic.

It is interesting in itself but also suggestive and inspiring in
the perspective of certain alternative models of $T$ to rewrite
(\ref{eq.194}) in other equivalent forms, namely,
\begin{eqnarray}
T_{\rm tr}&=&\frac{m}{2}g_{ij}v^{i}v^{j}=\frac{m}{2}G[\varphi]_{AB} \widehat{v}^{A}\widehat{v}^{B},\label{eq.196}\\
T_{\rm int}&=&\frac{1}{2}g_{ij}V^{i}{}_{A}V^{j}{}_{B}J^{AB}= 
\frac{1}{2}G[\varphi]_{KL}\widehat{\Omega}^{K}{}_{A} 
\widehat{\Omega}^{L}{}_{B}J^{AB}\nonumber\\
&=&\frac{1}{2}g_{ij}\Omega^{i}{}_{l}\Omega^{j}{}_{k}J[\varphi]^{kl},
\label{eq.197}
\end{eqnarray}
where ${\Omega}^{i}{}_{j}$, $\widehat{\Omega}^{K}{}_{L}$, $\widehat{v}^{K}$ are given respectively by (\ref{eq.130}), (\ref{eq.139}), and $J[\varphi]$ is the spatial, thus, $\varphi$-dependent, inertial tensor:
\begin{equation}
J[\varphi]^{kl}=\varphi^{k}{}_{A}\varphi^{l}{}_{B}J^{AB}. \label{eq.198}
\end{equation}

Let us remind that $J[\varphi]$ may be expressed in terms of the
configuration $\phi:N\rightarrow M$ as follows:
\begin{equation}
J[\phi]^{kl}=\int\left(y^{k}-x^{k}\right)\left(y^{l}-x^{l}\right)d\mu_{\phi}(y),
\label{eq.199}
\end{equation}
where $\mu_{\phi}$ is the $\phi$-transport of the measure $\mu$
from $N$ to $M$ (the Euler distribution of mass), and $x^{k}$ are
coordinates of the centre of mass in $M$. For affine
configurations (\ref{eq.199}) implies (\ref{eq.198}). 

If the internal inertia is isotropic, i.e., (\ref{eq.195}) holds, then (\ref{eq.197}) becomes the following geometry-based expression:
\begin{eqnarray}
T_{\rm int}&=&\frac{I}{2}g_{ij}V^{i}{}_{A}V^{j}{}_{B}\eta^{AB}= 
\frac{I}{2}G[\varphi]_{KL}\widehat{\Omega}^{K}{}_{A} 
\widehat{\Omega}^{L}{}_{B}\eta^{AB}\nonumber\\
&=&\frac{I}{2}g_{ij}\Omega^{i}{}_{l}\Omega^{j}{}_{k}C[\varphi]^{-1kl}.
\label{eq.197a}
\end{eqnarray}

The formula (\ref{eq.194}) works correctly in various problems. Its
characteristic feature is that it is not invariant under any
subgroup of ${\rm GL}(V)\times{\rm GL}(U)$ larger than 
${\rm O}\left(V,g\right)\times{\rm O}\left(U,J^{-1}\right)$. From the purely mathematical point of view this is some shortcoming, because on the primary kinematical level, internal degrees of freedom are ruled by the total ${\rm GL}(V)\times{\rm GL}(U)$. Thus, even the very curiosity motivated our earlier search of affinely-invariant models of the kinetic energy of a single affine
body \cite{acta,all04,all05}, or in more mathematical terms --- the search of
affinely-invariant Riemannian structures on the affine group. The
more so that afterwards some possibility of interesting physical
applications seemed to appear.

Let us remind the structure of affinely-invariant kinetic energies for a single affine body.

We begin with left-invariant models, i.e., ones invariant under
the spatial affine group ${\rm GAf}(M)$. This means they are invariant
under (\ref{eq.156}) assuming that $A$ runs over the total ${\rm GAf}(M)$,
and $B={\rm Id}_{N}$. If $T$ is to split additively into translational
and internal parts like in (\ref{eq.194}), then the only natural and
at the same time maximally general nonrelativistic models have the
following form:
\begin{equation}
T=T_{\rm tr}+T_{\rm int}=\frac{m}{2}\eta_{AB}\widehat{V}^{A}\widehat{V}^{B}+ \frac{1}{2}\mathcal{L}^{B}{}_{A}{}^{D}{}_{C}\widehat{\Omega}^{A}{}_{B} \widehat{\Omega}^{C}{}_{D},\label{eq.200}
\end{equation}
where $\mathcal{L}^{B}{}_{A}{}^{D}{}_{C}$ are constants.
Obviously, they are symmetric in bi-indices:
\begin{equation}
\mathcal{L}^{B}{}_{A}{}^{D}{}_{C}=\mathcal{L}^{D}{}_{C}{}^{B}{}_{A}. \label{eq.201}
\end{equation}

Let us observe that such kinetic energies, i.e., metric tensors on
the configuration space $Q=M\times{\rm LI}\left(U,V\right)$ are affinely
invariant in $M$, but not in the material space $N$. Namely,
(\ref{eq.200}) is invariant only under such material endomorphisms,
i.e., elements of ${\rm GL}(U)$, which preserve the material tensors
$\varphi\in U^{\ast}\otimes U^{\ast}$, $\mathcal{L}\in U \otimes
U^{\ast}\otimes U\otimes U^{\ast}$. The highest possible symmetry in
$N$ is $\eta$-rotational one, i.e., ${\rm O}\left(U,\eta\right)\subset{\rm GL}(U)$. It is attained when $\mathcal{L}$ is algebraically built of $\eta$ and
${\rm Id}_{U}$, i.e.,
\begin{equation}
\mathcal{L}^{B}{}_{A}{}^{D}{}_{C}=I\eta^{BD}\eta_{AC}+ 
A\delta^{B}{}_{C}\delta^{D}{}_{A}+B\delta^{B}{}_{A}\delta^{D}{}_{C}. \label{eq.202}
\end{equation}
Then $T_{\rm int}$ in (\ref{eq.200}) may be concisely written as
\begin{equation}
T_{\rm int}=\frac{I}{2}{\rm Tr}\left(\widehat{\Omega}^{T\eta}
\widehat{\Omega}\right)+\frac{A}{2}{\rm Tr}\left(\widehat{\Omega}^{2}\right)+ \frac{B}{2}\left({\rm Tr}\; \widehat{\Omega}\right)^{2}, \label{eq.203}
\end{equation}
where $\widehat{\Omega}^{T\eta}$ denotes the transpose of the mixed tensor $\widehat{\Omega}$ performed in the $\eta$-sense:
\begin{equation}
\left(\widehat{\Omega}^{T\eta}\right)^{A}{}_{B}= 
\eta^{AC}\eta_{BD}\widehat{\Omega}^{D}{}_{C}. \label{eq.204}
\end{equation}
This is the "usual" transpose when orthonormal coordinates are used in $U$ and $\eta_{AB}=_{\ast}\delta_{AB}$. Obviously, the inertial coefficients $I$, $A$, $B$ are constants.

The metric tensor $\Gamma$ underlying (\ref{eq.200}) has the following form:
\begin{equation}
\Gamma=mC[\varphi]_{ij}dx^{i}\otimes dx^{j}+ 
\frac{1}{2}\mathcal{L}^{A}{}_{B}{}^{C}{}_{D}\varphi^{-1B}{}_{i} \varphi^{-1D}{}_{j}d\varphi^{i}{}_{A}\otimes\varphi^{j}{}_{B}. \label{eq.205}
\end{equation}
It is evidently curved, $\left(Q,\Gamma\right)$ is a Riemann space with the non-vanishing curvature tensor. Compare this with the flat, Euclidean structure of (\ref{eq.196}), (\ref{eq.197}), where the metric tensor of $T$ was given by
\begin{equation}
\Gamma=mg_{ij}dx^{i}\otimes dx^{j}+
g_{ij}J^{AB}d\varphi^{i}{}_{A}\otimes\varphi^{j}{}_{B}. \label{eq.206}
\end{equation}

Let us observe that the last two terms in (\ref{eq.203}) are affinely invariant both in the physical and material spaces. This is impossible for the total kinetic energy, when translational motion is taken into account. The combination of the mentioned $A,B$-terms is not positively definite, for example, the $A$-term, $\left(A/2\right){\rm Tr}\left(\widehat{\Omega}^{2}\right)$, has the following signature:
\begin{equation}
{\rm sign}\ A\left(\frac{n(n-1)}{2}\; -,\frac{n(n+1)}{2}\; +\right). 
\end{equation}
Roughly speaking, the negative and positive contributions (if $A>0$) correspond respectively to the "compact and non-compact directions" in the linear group. The negative contributions to $T_{\rm int}$ might seem physically embarrassing, nevertheless, they are not only harmless, but even may be desirable in certain dynamical models. Incidentally $T_{\rm int}$ is positively definite in certain open domain of inertial parameters $\left(I,A,B\right)\in\mathbb{R}^{3}$.

Let us observe that the model (\ref{eq.200}) specialized to the particular case (\ref{eq.202}), (\ref{eq.203}) may be formally "obtained" from the combination of (\ref{eq.196}), (\ref{eq.197}), (\ref{eq.197a}) by substituting the constant material metric $\eta$ instead of the $\varphi$-dependent Green tensor $G[\varphi]$. This was just the heuristic motivation for rewriting (\ref{eq.194}) in the form (\ref{eq.196}), (\ref{eq.197}). One obtains then some 
quadratic function of non-holonomic Lie-algebraic velocities $\left(\widehat{v}^{A},\widehat{\Omega}^{A}{}_{B}\right)$ with constant coefficients, just as (\ref{eq.194}) is a constant-coefficients quadratic form of holonomic generalized velocities $\left(v^{i},V^{i}{} _{A}\right)$. This seemingly formal modification is, as a matter of fact, very essential because it leads to affinely-invariant (in $M$) kinetic energies, i.e., to $M$-affinely metric tensors on the configuration space.

Repeating the above way of thinking in terms of Lie-algebraic variables $\left(v^{i},\Omega^{i}{}_{j}\right)$ one finds the heuristic way towards models affine-invariant in the material space $N$, i.e., ${\rm GAff}(N)$-invariant ones. The only natural ${\rm GAff}(N)$-counterpart of (\ref{eq.205}), without interference between translational and internal parts has the form:
\begin{equation}\label{eq.207}
T=T_{\rm tr}+T_{\rm int}=\frac{m}{2}g_{ij}v^{i}v^{j}+
\frac{1}{2}\mathcal{R}^{j}{}_{i}{}^{l}{}_{k}\Omega^{i}{}_{j}\Omega^{k}{}_{l},
\end{equation}
with symmetry properties of $\mathcal{R}$ like those of $\mathcal{L}$ in (\ref{eq.201}). Dually to (\ref{eq.200}), $T_{\rm tr}$ is not invariant under ${\rm GAff}(M)$, and in a consequence of the structure of the affine group, there is no doubly invariant metric on ${\rm GAff}(n,\mathbb{R})$. However, in analogy to (\ref{eq.202}), there are models invariant under ${\rm GAff}(N)$ and simultaneously under the Euclidean group ${\rm E}(M,g)$, thus, in particular, under the orthogonal group ${\rm O}(V,g)$. The corresponding formula, dual to (\ref{eq.200}), (\ref{eq.203}), is
\begin{equation}\label{eq.208}
T_{\rm int}=\frac{I}{2}{\rm Tr}\left(\Omega^{Tg}\Omega\right)+
\frac{A}{2}{\rm Tr}\left(\Omega^{2}\right)+
\frac{B}{2}\left({\rm Tr}\; \Omega\right)^{2},
\end{equation}
where, obviously,
\begin{equation}\label{eq.209}
\left(\Omega^{Tg}\right)^{i}{}_{j}=g^{ik}g_{jl}\Omega^{l}{}_{k},\qquad
{\rm Tr}\left(\Omega^{2}\right)={\rm Tr}\left(\widehat{\Omega}^{2}\right),\qquad 
{\rm Tr}\; \Omega={\rm Tr}\; \widehat{\Omega}.
\end{equation}
This means that
\begin{equation}\label{eq.210}
\mathcal{R}^{j}{}_{i}{}^{l}{}_{k}=Ig^{jl}g_{ik}+
A\delta^{j}{}_{k}\delta^{l}{}_{i}+
B\delta^{j}{}_{i}\delta^{l}{}_{k}.
\end{equation}
Unlike the total kinetic energy, the internal one $T_{\rm int}$ may be simultaneously affinely invariant in $M$ and $N$ and is then equal to the sum of the last two terms in (\ref{eq.203}) or (\ref{eq.208}), they are equal to each other.

It is very instructive to see what would be the most general kinetic energy, i.e., metric tensor on the configuration space, invariant under the isometry groups in $M$ and $N$, ordering its terms according to the additional "increasing degree" of the affine invariance in space and in the body.

The corresponding kinetic energy may be written down as follows:
\begin{eqnarray}\label{eq.211}
T&=&\frac{1}{2}\left(m_{1}G_{AB}+m_{2}\eta_{AB}\right)
\widehat{v}^{A}\widehat{v}^{B}+
\frac{A}{2}{\rm Tr}\left(\widehat{\Omega}^{2}\right)+
\frac{B}{2}\left({\rm Tr}\; \widehat{\Omega}\right)^{2}\\
&+&\frac{1}{2}\left(I_{1}G_{KL}G^{MN}+
I_{2}\eta_{KL}\eta^{MN}+I_{3}G_{KL}\eta^{MN}+I_{4}\eta_{KL}G^{MN}
\right)\widehat{\Omega}^{K}{}_{M}\widehat{\Omega}^{L}{}_{N},\nonumber
\end{eqnarray}
what may be also written in the following alternative form:
\begin{eqnarray}\label{eq.212}
T&=&\frac{1}{2}\left(m_{1}g_{ij}+m_{2}C_{ij}\right)v^{i}v^{j}+
\frac{A}{2}{\rm Tr}\left(\Omega^{2}\right)+\frac{B}{2}\left({\rm Tr}\; \Omega\right)^{2}\\
&+&\frac{1}{2}\left(I_{1}g_{kl}g^{mn}+
I_{2}C_{kl}C^{mn}+I_{3}g_{kl}C^{mn}+I_{4}C_{kl}g^{mn}
\right)\Omega^{k}{}_{m}\Omega^{l}{}_{n}.\nonumber
\end{eqnarray}
{\bf Remark:} $G^{KL}$ are not "$\eta$-raised" $G_{AB}$, but their inverses; similarly, $C^{ij}$ are contravariant inverses of $C_{ab}$, not their contravariant "$g$-raised" versions.

It is very important to say what is the metrical-affine interplay in the expression (\ref{eq.211}) or (\ref{eq.212}). If we put $m_{1}$, $I_{1}$, $I_{3}$, $I_{4}$, we obtain the model affinely invariant in $M$ but only $\eta$-metrically (and not higher) invariant in $N$. Models $g$-metrically invariant in $M$ (but not higher invariant there) and affinely invariant in $N$ are obtained when we choose $m_{2}$, $I_{2}$, $I_{3}$, $I_{4}$. Or, more correctly, there is also something between "metrical" and "affine". And this also has to do with the remark a few lines above. Namely, admitting in (\ref{eq.211}), (\ref{eq.212}) the term controlled by $I_{2}$, we obtain the model in which the internal kinetic energy is invariant under both $\eta$-rotations and dilatations in the material space, thus, under the group of similarities (linear-conformal group $\mathbb{R}^{+}{\rm O}(U,\eta)$). Similarly, the $I_{1}$-controlled term is invariant under $\mathbb{R}^{+}{\rm O}(V,g)$, thus, also under dilatations in the physical space. Of course, we mean here the linear-conformal invariance of the internal term alone; it does not seem possible to unify the total affine invariance in $N$ with the rotational-dilatational total invariance in $M$, and conversely. The translational term $T_{\rm tr}$ is an obstacle against such a joint invariance.

Obviously, the metric tensor underlying (\ref{eq.211}), (\ref{eq.212})
\begin{equation}\label{eq.213}
\mathcal{G}=\mathcal{G}({\rm tr})_{ij}dx^{i}\otimes dx^{j}+ 
\mathcal{G}({\rm int})^{A}{}_{k}{}^{B}{}_{l}d\varphi^{k}{}_{A}\otimes d\varphi^{l}{}_{B},
\end{equation}
where
\begin{eqnarray}
\mathcal{G}({\rm tr})_{ij}&=&m_{1}g_{ij}+m_{2}C_{ij},\label{eq.214}\\
\mathcal{G}({\rm int})^{A}{}_{k}{}^{B}{}_{l}&=&I_{1}g_{kl}G^{AB}+
I_{2}C_{kl}\eta^{AB}+I_{3}g_{kl}\eta^{AB}+I_{4}C_{kl}G^{AB}\nonumber\\
&+&A\varphi^{-1A}{}_{l}\varphi^{-1B}{}_{k}+
B\varphi^{-1A}{}_{k}\varphi^{-1B}{}_{l}.\label{eq.215}
\end{eqnarray}
Let us stress again that $\mathcal{G}({\rm int})$ is autonomous in the sense that its coefficients depend only on $\varphi$ but not on $x$. Unlike this, $\mathcal{G}({\rm tr})$ is non-autonomous in the sense that it depends on internal variables $\varphi$ (if $m_{2}\neq 0$).

It may be also interesting to express (\ref{eq.214}), (\ref{eq.215}) in various non-holonomic representations,
\begin{eqnarray}
\mathcal{G}({\rm tr})&=&\mathcal{G}({\rm tr})_{AB}\widehat{\omega}^{A}\otimes\widehat{\omega}^{B},\label{eq.216a}\\
\mathcal{G}({\rm tr})_{AB}&=&m_{1}G_{AB}+
m_{2}\eta_{AB},\quad \widehat{\omega}^{A}=\varphi^{-1A}{}_{i}dx^{i}=\varphi^{-1A}{}_{i}\omega^{i},
\label{eq.216b}\\
\mathcal{G}({\rm int})&=&\mathcal{G}({\rm int})^{m}{}_{k}{}^{n}{}_{l}
\omega^{k}{}_{m}\otimes\omega^{l}{}_{n}=
\mathcal{G}({\rm int})^{B}{}_{A}{}^{D}{}_{C}
\widehat{\omega}^{A}{}_{B}\otimes\widehat{\omega}^{C}{}_{D},
\label{eq.217a}\\
\mathcal{G}({\rm int})^{m}{}_{k}{}^{n}{}_{l}&=&
I_{1}g_{kl}g^{mn}+I_{2}C_{kl}C^{mn}+I_{3}g_{kl}C^{mn}+I_{4}C_{kl}g^{mn}
\nonumber\\
&+&A\delta^{m}{}_{l}\delta^{n}{}_{k}+B\delta^{m}{}_{k}\delta^{n}{}_{l},
\label{eq.217b}\\
\mathcal{G}({\rm int})^{B}{}_{A}{}^{D}{}_{C}&=&
I_{1}G_{AC}G^{BD}+I_{2}\eta_{AC}\eta^{BD}+I_{3}G_{AC}\eta^{BD}+
I_{4}\eta_{AC}G^{BD}\nonumber\\
&+&A\delta^{B}{}_{C}\delta^{D}{}_{A}+B\delta^{B}{}_{A}\delta^{D}{}_{C},
\label{eq.217c}\\
\omega^{a}{}_{b}&=&\varphi^{-1K}{}_{b}d\varphi^{a}{}_{K},\qquad \widehat{\omega}^{A}{}_{B}=\varphi^{-1A}{}_{i}d\varphi^{i}{}_{B}.
\label{eq.217d}
\end{eqnarray}
Roughly speaking, these are metric tensors on the affine group ${\rm GAff}(n,\mathbb{R})$ invariant under the left (spatial) and right (material) action of isometries. By appropriate specification of constants we obtain metrics on ${\rm GAff}(n,\mathbb{R})$ with larger isometry groups, containing left or right action of affine transformations.

This is mathematics, differential geometry, interesting in itself. But there are also some physical ideas and hypotheses beyond. Namely, in condensed matter it is quite reasonable to expect that the effective inertial properties of molecules are effective, collective resultants of the interaction of any particle with its surrounding, instead of being determined by the "true" metric tensor of space. There are well-known physical examples in solid state physics, namely, the effective mass tensor of the electrons in crystals.

Our guiding idea is that the search of such effective multiparticle models should be based rather on some natural symmetry postulates than on the systematic "derivation" from some more "microscopic" structural assumptions and calculations, like, e.g., ones on the "atomic" level. In view of the very complicated system of equations principally describing the problem on the level of electrons-nuclei systems, the latter task would be completely hopeless. In such situations it happens often that symmetry principles are the only ones which enable us to obtain some qualitative and sometimes also quantitative results.

On the non-dissipative Hamiltonian level, our dynamics will be based mainly on the Hamilton functions of the potential shape:
\begin{equation}\label{eq.221}
H=\mathcal{T}+\mathcal{V},
\end{equation}
where $\mathcal{T}$ is obtained from the kinetic energy models $T$ (metrics on the configuration space) based on the additive postulate (\ref{eq.190}), where, for any fixed particle $\mathcal{K}$, $T(\mathcal{K})$ has one of the postulated forms (\ref{eq.194}), (\ref{eq.200}), (\ref{eq.203}), (\ref{eq.207}), (\ref{eq.208}), (\ref{eq.211}), (\ref{eq.212}). $\mathcal{T}$ is obtained from $T$ in a standard way via the Legendre transformation. Usually it is convenient to express this transformation in terms of non-holonomic velocities like $\Omega^{i}{}_{j}$, $\widehat{\Omega}^{A}{}_{B}$, $\widehat{v}^{A}$, because then $\mathcal{T}$ is represented as a function of affine spin in the spatial or co-moving description $\Sigma^{i}{}_{j}$, $\widehat{\Sigma}^{A}{}_{B}$ and of the linear momentum, also in both possible representations, $p_{i}$, $\widehat{p}_{A}$. This is a very effective procedure, because equations of motion may be obtained almost directly from (\ref{eq.185}) by substituting as phase-space functions $F$ the mentioned Hamiltonian generators, i.e., momentum mappings, e.g., $\left(p_{i},\Sigma^{i}{}_{j}\right)$ or $\left(\widehat{p}_{A},\widehat{\Sigma}^{A}{}_{B}\right)$. The point is that Poisson brackets of these functions are well known, given by structure constants of the affine group ${\rm GAff}(n,\mathbb{R})$.

For simple models, e.g., (\ref{eq.203}), (\ref{eq.208}), the Legendre transformation may be explicitly inverted and the first-order Hamilton-type equations (\ref{eq.164}) may be easily expressed as second-order differential equations for the time-dependence of generalized coordinates $x^{i}$, $\varphi^{i}{}_{A}$ for the single affine body, and for $x^{i}(\mathcal{K})$, $\varphi^{i}{}_{A}(\mathcal{K})$ for the system of affine bodies.

Clearly, Poisson brackets for quantities concerning different "particles", $\mathcal{K}\neq\mathcal{L}$, are always vanishing, because they are functions of different generalized coordinates and momenta. This simplifies remarkably derivation of equations of motion.

\section*{Summary}

Let us summarize the ideas and programme outlined here. Discussed was hierarchy of dynamical quantities describing systems of affine bodies. In Section 2 we reviewed scalar invariants for pairs of affine bodies. The mentioned hierarchy concerned the degree of affine invariance. We discussed systems of scalars invariant under isometries and under the total affine group in physical and material spaces. Quite different problem, to some extent experimental one, is whether really the affine invariants may be useful in physical models of systems of mutually interaction affine bodies. There is also a stronger question: Perhaps in highly condensed matter the purely affine scalars might be sufficient for constructing viable dynamical models of mutual interactions (including both translational and internal degrees of freedom). Is it so or not? There are some physical reasons to expect such a possibility. From the purely geometric point of view of rational mechanics the problem of such Thales-like physics is interesting in itself. Our two-body affine and metrical scalars constructed in Section 2 are thought on as argument of binary potentials describing interactions in systems of affine bodies ("molecules"). Let us repeat: affinely invariant scalars do exist only for systems of affine bodies, not for the single affine body. Affinely-invariant kinetic energies (metrics on the configuration space) are well defined both for single affine bodies and for their systems. For the systems we have the additive formulas (\ref{eq.190}), or perhaps the binary ones (\ref{eq.191}), when something like the gyroscopic coupling of internal (angular or affine) velocities occurs. As single kinetic energies, the formulas like (\ref{eq.197}), (\ref{eq.200}), (\ref{eq.203}), (\ref{eq.207}), (\ref{eq.208}), (\ref{eq.211}), (\ref{eq.212}), etc.\ may be chose, depending on the postulated models, i.e., on the assumed metric tensor of the configuration space. Binary expressions $T(\mathcal{K},\mathcal{L})$ are obtained as so-called polarizations of the one-particle quadratic forms $T(\mathcal{K})$. The underlying geometry, physical motivation and expected usefulness of the particular models of $T(\mathcal{K})$ were discussed in our earlier papers \cite{acta,all04,all05}.

The "traditional" model (\ref{eq.194}) is derived for the usual affinely constrained system of material points. It is also compatible with the d'Alembert principle meant in the sense of the usual Euclidean metric $g$. In this sense it seems to be the most physical model. Nevertheless, there are also some other models, like (\ref{eq.200})--(\ref{eq.211}). They are geometrically more interesting, because they have "large" symmetry groups and the resulting essential nonlinearity in the kinetic, inertial part of Lagrangian. And the point is that they seem to offer some unexpected dynamical applications. Namely, even on the purely geodetic level, without any potential, the doubly affinely invariant models of internal kinetic energy, i.e., (\ref{eq.203}), (\ref{eq.208}) without the metrical term, i.e., with $I=0$, may encode the dynamics of nonlinear elastic vibrations. The general solution contains an open family of bounded orbits, even if no potential is assumed. In any case, it is so, when the motion is isochoric, i.e., the body is incompressible. For the compressible case, some purely dilatational potential may be used to stabilized dilatations. And it is interesting that the main characteristics of motion my be obtained on the basis of exponential functions of matrices. In such bounded solutions, the deformation invariants perform finite vibrations. In multibody systems this concerns both the usual deformation invariants and invariants of mutual deformation tensors defined in Section 2. The peculiarity of multibody affine systems is that one can also use affinely-invariant potential energies of binary interactions. This offers a wide class of dynamical models to be postulated as a hypothetical description of interactions in discrete structured media. Their continuous limit will lead to a variety of dynamical models of structured continua.

When dealing with objects like molecules, fullerens, graphens and clusters, one must use the quantized version of the model. Certain primary ideas in this direction were presented in \cite{JJS_05,JJS_2}.

\section*{Acknowledgements}

This paper partially contains results obtained within the framework of the research project 501 018 32/1992 financed from the Scientific Research Support Fund in 2007-2010. The authors are greatly indebted to the Ministry of Science and Higher Education for this financial support. The support within the framework of Institute internal programme 203 is also greatly acknowledged.

We are also greatly indebted to Professor P. M. Mariano for discussions with him and for encouraging us to the work in this direction.

\end{document}